%% file: n-PrCN-vib_07.tex
  \documentclass[structabstract]{aa}
\usepackage{graphicx}
\usepackage{txfonts}
\usepackage{natbib}
\usepackage{ulem}

\bibpunct[]{(}{)}{;}{a}{}{,}

\begin{document}
   \title{Laboratory spectroscopic study and astronomical detection of 
          vibrationally excited \textit{n}-propyl cyanide}

   \author{Holger S.~P. M{\"u}ller\inst{1}
           \and
           Adam Walters\inst{2,3}
           \and
           Nadine Wehres\inst{1}
           \and
           Arnaud Belloche\inst{4}
           \and
           Olivia H. Wilkins\inst{1}
           \and
           Delong Liu\inst{2,3}
           \and
           R{\'e}mi Vicente\inst{2,3}
           \and
           Robin T. Garrod\inst{5}
           \and
           Karl M. Menten\inst{4}
           \and
           Frank Lewen\inst{1}
           \and
           Stephan Schlemmer\inst{1}
           }

   \institute{I.~Physikalisches Institut, Universit{\"a}t zu K{\"o}ln,
              Z{\"u}lpicher Str. 77, 50937 K{\"o}ln, Germany\\
              \email{hspm@ph1.uni-koeln.de}
              \and
              Universit{\'e} de Toulouse 3, OMP, IRAP, Toulouse, France
              \and
              CNRS, IRAP, 9 Av. Colonel Roche, BP 44346, 31028 Toulouse cedex 4, France
              \and
              Max-Planck-Institut f\"ur Radioastronomie, Auf dem H\"ugel 69, 
              53121 Bonn, Germany
              \and
              Departments of Chemistry and Astronomy, University of Virginia, Charlottesville, 
              VA 22904, USA
              }

   \date{Received 14 July 2016 / Accepted 23 August 2016}

  \abstract
{We performed a spectral line survey called Exploring Molecular Complexity with ALMA 
 (EMoCA) toward Sagittarius~B2(N) between 84.1 and 114.4~GHz with the Atacama Large 
 Millimeter/submillimeter Array (ALMA) in its Cycles~0 and 1. We determined line intensities  
 of \textit{n}-propyl cyanide in the ground vibrational states of its \textit{gauche} and 
 \textit{anti} conformers toward the hot molecular core Sagittarius~B2(N2) which suggest 
 that we should also be able to detect transitions pertaining to excited vibrational states.}
{We wanted to determine spectroscopic parameters of low-lying vibrational states of both 
 conformers of \textit{n}-propyl cyanide to search for them in our ALMA data.}
{We recorded laboratory rotational spectra of \textit{n}-propyl cyanide in two spectral 
 windows between 36 and 127~GHz. We searched for emission lines produced by these states 
 in the ALMA spectrum of  Sagittarius~B2(N2). We modeled their emission and the emission of 
 the ground vibrational states assuming local thermodynamic equilibrium (LTE).}
{We have made extensive assignments of $a$- and $b$-type transitions of the four lowest 
 vibrational states of the \textit{gauche} conformer which reach $J$ and $K_a$ quantum 
 numbers of 65 and 20, respectively. We assigned mostly $a$-type transitions for the 
 \textit{anti} conformer with $J$ and $K_a$ quantum numbers up to 48 and 24, respectively. 
 Rotational and Fermi perturbations between two \textit{anti} states allowed us to determine 
 their energy difference. The resulting spectroscopic parameters enabled us to identify 
 transitions of all four vibrational states of each conformer in our ALMA data. The emission 
 features of all states, including the ground vibrational state, are well-reproduced with the 
 same LTE modeling parameters, which gives us confidence in the reliability of the 
 identifications, even for the states with only one clearly detected line.}
{Emission features pertaining to the highest excited vibrational states of \textit{n}-propyl 
 cyanide reported in this work have been identified just barely in our present ALMA data. 
 Features of even higher excited vibrational states may become observable in future, more 
 sensitive ALMA spectra to the extent that the confusion limit will not have been reached. 
 The $^{13}$C isotopomers of this molecule are expected to be near the noise floor of our 
 present ALMA data. We estimate that transitions of vibrationally excited \textit{iso}-propyl 
 cyanide or aminoacetonitrile, for example, are near the noise floor of our current data 
 as well.}

\authorrunning{H.~S.~P. M{\"u}ller et al.}

\maketitle
\hyphenation{For-schungs-ge-mein-schaft}

\section{Introduction}
\label{intro}

Radio astronomical observations of the warm (> 100~K) and dense parts of star-forming 
regions known as hot cores or hot corinos have revealed a wealth of complex molecules. 
These are, for the most part, saturated or nearly saturated with up to 12 atoms thus far 
\citep{i-PrCN_det_2014}. Details can be found on the Molecules in Space web page 
(http://www.astro.uni-koeln.de/cdms/molecules) of the Cologne Database for 
Molecular Spectroscopy, CDMS, \citep{CDMS_1,CDMS_2} or on the list of Interstellar 
\& Circumstellar Molecules (http://www.astrochymist.org/astrochymist\_ism.html) of 
The Astrochymist web page (http://www.astrochymist.org/). A large portion of these 
molecules were detected toward Sagittarius (Sgr for short) B2(N). Sgr~B2(N) is part 
of the Sgr~B2 molecular cloud complex, one of the most prominent star-forming regions 
in our Galaxy and close to its center. There are two major sites of high-mass star 
formation, Sgr~B2(M) and Sgr~B2(N), see, for example, \citet{Sgr-B2_structure_2016}, 
of which Sgr~B2(N) has a greater variety of complex organic molecules \citep{SgrB2_3mm_2013}. 
It contains two dense compact hot cores, the more prominent Sgr~B2(N1), also known as 
Large Molecule Heimat, and Sgr~B2(N2), about $5''$ to the north of it 
\citep{deuterated_SgrB2N2_2016,Sgr-B2_structure_2016}.

Following up on our single-dish survey of the 3~mm wavelength range with the IRAM 30~m 
telescope \citep{SgrB2_3mm_2013}, we have used the Atacama Large Millimeter/submillimeter Array 
(ALMA) in its Cycles~0 and 1 to perform a spectral line survey of Sgr~B2(N) between 84.1 and 
114.4~GHz \citep{deuterated_SgrB2N2_2016}. The name of the survey, Exploring Molecular 
Complexity with ALMA (EMoCA), describes the main motivation of this study. The high angular 
resolution ($\sim$1.6$''$) allows us to separate the emission of the two hot cores, that, 
in turn, revealed the narrow line width of Sgr~B2(N2) ($\sim$5~km\,s$^{-1}$), greatly reducing 
the line confusion with respect to our previous single-dish survey of Sgr~B2(N). 
As a consequence, we have focused our analysis on Sgr~B2(N2) thus far.

The detection of \textit{iso}-propyl cyanide  (or \textit{i}-PrCN for short; 
\textit{i}-C$_3$H$_7$CN) as the first branched alkyl molecule in space was one of the first 
results of EMoCA \citep{i-PrCN_det_2014}. The laboratory spectroscopic investigation on 
\textit{i}-PrCN \citep{i-PrCN_rot_2011} was an obvious prerequisite, but the decrease in 
line confusion was important also because this molecule was not detected in our previous 
single-dish survey of Sgr~B2(N) \citep{SgrB2_3mm_2013,i-PrCN_rot_2011}. Interestingly, 
\textit{i}-PrCN is not much less abundant than its straight-chain isomer \textit{normal}-propyl 
cyanide (\textit{n}-PrCN for short) \citep{i-PrCN_det_2014}. Other published results of EMoCA 
include the detection of the three isotopomers of ethyl cyanide with two $^{13}$C 
\citep{EtCN_2x13C_2016}, an extensive study of deuterated complex organic molecules in 
Sgr~B2(N2) with, for example, the first, albeit tentative, detection of deuterated ethyl cyanide 
\citep{deuterated_SgrB2N2_2016}, and a study of alkanols and alkanethiols, which yielded a 
rather stringent upper limit to the column density of ethanethiol in this source 
\citep{RSH_ROH_SgrB2N2_2016}.

We had detected \textit{n}-PrCN initially in our single-dish study \citep{det-PrCN_EtFo}. 
The line intensities of this molecule in our ALMA spectra toward Sgr~B2(N2) in combination 
with its several low-lying vibrational states 
\citep{n-prCN_conf_IR_1976,n-prCN_conf_IR_1987,n-PrCN_IR_etc_2001} suggested that we should 
be able to detect transitions of vibrationally excited \textit{n}-PrCN in our ALMA data.


\begin{figure} 
\centering 
\includegraphics[angle=0,width=4.0cm]{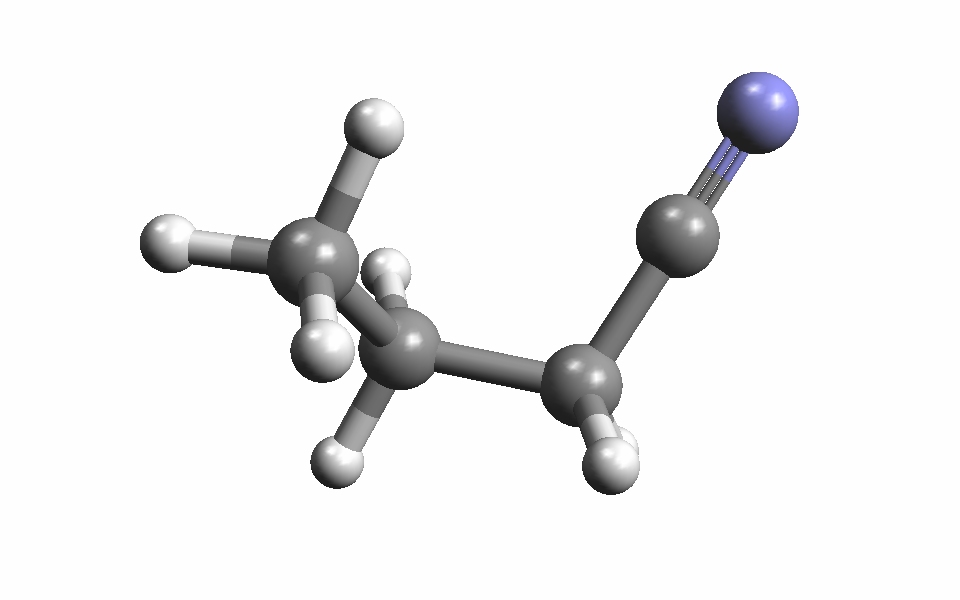} \includegraphics[angle=0,width=4.0cm]{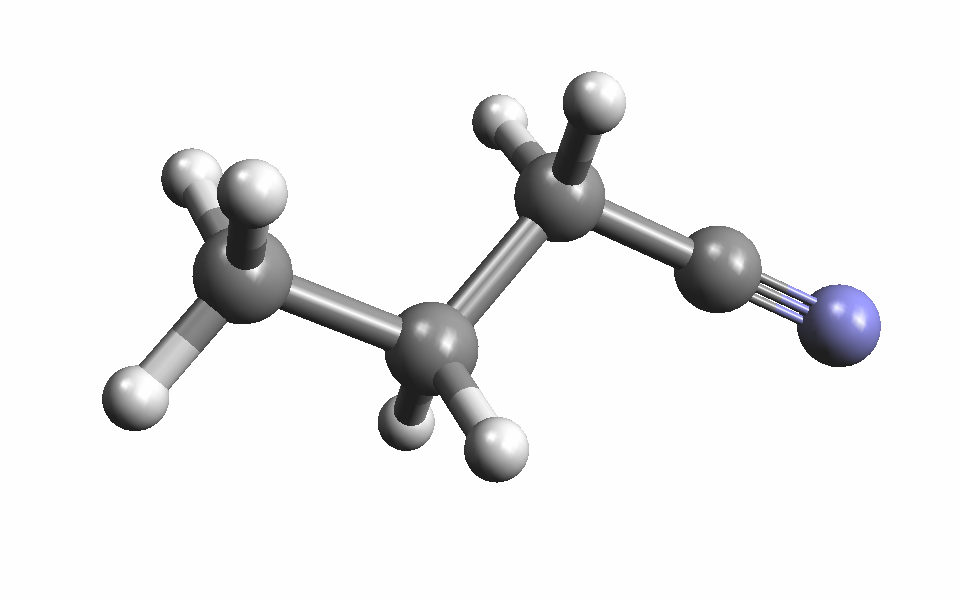} 
  \caption{Schematic depiction of the \textit{gauche} (left) and \textit{anti} (right) conformers of 
           \textit{normal}-propyl cyanide. The C and N atoms are represented by gray and violet 
           ``spheres'', respectively, and the H atoms by small, light gray ones.}
\label{molecule_models} 
\end{figure} 


The ground state rotational spectrum has been investigated quite well in the laboratory. 
\citet{n-PrCN_rot_1962} used microwave spectroscopy between 8 and 32~GHz to study rotational 
isomerism (i.e., the presence and energetics of different conformers) of \textit{n}-PrCN. 
He identified two 
possible conformers, \textit{anti} (\textit{a} for short, also known as 
\textit{trans}) with a dihedral CCCC angle of 180$^{\circ}$ and \textit{gauche} (\textit{g} for 
short, also known as \textit{synclinal}) with a dihedral CCCC angle of about $\pm$60$^{\circ}$, 
see also Fig.~\ref{molecule_models}. The two \textit{gauche} conformers are indistinguishable. 
\citet{n-PrCN_rot_1982} studied the $^{14}$N quadrupole structure of the \textit{anti} conformer 
by Fourier transform microwave spectroscopy. \citet{n-PrCN_FTMW_1988} used the same technique 
between 4.8 and 26.4~GHz for an extensive study of the rotational spectrum, quadrupole structure, 
and methyl internal rotation of both conformers. \citet{n-PrCN_mmW_1988} extended the measurements 
into the millimeter region, measured the dipole moment components, and determined from intensity 
measurements that the \textit{anti} conformer is lower in energy than \textit{gauche} by 
$1.1 \pm 0.3$~kJ~mol$^{-1}$ (or $92 \pm 25$~cm$^{-1}$ or $132 \pm 36$~K), in line with 
the estimate by \citet{n-PrCN_rot_1962} that the energy difference between the two is small, 
perhaps less than 1~kcal~mol$^{-1}$ (or 350~cm$^{-1}$ or 504~K). However, 
\citet{n-prCN_conf_IR_1976} and \citet{n-PrCN_e-diff_2000} determined from earlier infrared 
(IR) spectroscopy of solid, liquid, and gaseous samples and from electron diffraction of a 
gaseous sample that the \textit{gauche} conformer is lower in energy. \citet{n-PrCN_IR_etc_2001} 
used IR spectroscopy of \textit{n}-PrCN dissolved in liquid Xenon ($-$60 to $-100^{\circ}$C) 
to determine that the \textit{gauche} conformer is lower than \textit{anti} by 
$0.48 \pm 0.04$~kJ~mol$^{-1}$ (or $40 \pm 3$~cm$^{-1}$ or $58 \pm 4$~K). Using our ALMA data, 
we modeled the \textit{anti}/\textit{gauche} energy difference and found it to be fully 
consistent with the latter value \citep{i-PrCN_det_2014}.

The experimental data on excited vibrational states were rather limited. \citet{n-PrCN_rot_1962} 
also observed vibrational satellites of \textit{n}-PrCN in his microwave study and published 
rotational constants for the four lowest states of each conformer. They belong in either case 
to the three lowest fundamental states as well as to the overtone state of the lowest one. 
These parameters turned out to be sufficient for assignments of laboratory spectra near 1~cm 
wavelength, but were by far too imprecise for assignments in our ALMA data.

Therefore, we recorded rotational spectra of \textit{n}-PrCN in two spectral windows, 36 to 
70~GHz and 89 to 127~GHz, and analyzed the four lowest vibrational states of each conformer. 
These results enabled us to detect transitions of all four states in our ALMA data. Details 
on our laboratory spectroscopic investigations are given in Sect.~\ref{exptl}. Background 
information on the rotational and vibrational spectroscopy of \textit{n}-PrCN is provided 
in Sect.~\ref{rot_vib_backgr}. Results of our current laboratory spectroscopic study and 
a related discussion are given in Sect.~\ref{lab-results}. Section~\ref{astro-results} 
describes the analyses of our ALMA data, and Sect.~\ref{conclusion} contains conclusions 
and an outlook.

\section{Laboratory spectroscopic details}
\label{exptl}

All measurements were carried out at room temperature at the Universit{\"a}t zu K{\"o}ln 
in two connected 7~m long double path absorption cells with inner diameter of 100~mm which 
were equipped with Teflon windows. A commercial sample of \textit{n}-PrCN was flowed slowly 
through the cell at pressures of around 1~Pa. We used frequency modulation throughout with 
demodulation at $2f$, which causes an isolated line to appear approximately as a second 
derivative of a Gaussian.

The entire region between 36 and 70~GHz was covered in successive scans with an Agilent E8257D 
microwave synthesizer as source and a Schottky diode detector. The point spacing was 20~kHz, 
and at each point, data were accumulated for 20~ms. One upward scan and one downward scan 
were coadded after completion of the spectral recording. Later, we recorded smaller sections 
of 2 to 6~MHz widths around the predicted positions of $b$-type transition of the \textit{anti} 
conformer employing much longer overall integration times which reached up to 50~min for 6~MHz. 
The entire region between 89.25 and 126.75~GHz was covered similarly using a Virginia Diode, 
Inc. (VDI) tripler driven by a Rohde \& Schwarz SMF~100A synthesizer as source and a Schottky 
diode detector. The point spacing in this frequency region was 60~kHz, and the integration 
time per point was again 20~ms with scans in both directions Further details on the spectrometer 
system are available elsewhere \citep{1-2-PD_rot_2014,D-EtOH_rot_2015}.

Despite the long integration times, the confusion limit was not reached by far in either of 
these two large spectral windows. The sensitivity of the spectrometer systems varied greatly 
with frequency. It was very low at the lower ends of both frequency windows and near 127~GHz. 
In addition, there were sections in which the sensitivity was very low within the spectral windows. 
In addition, intensities sometimes changed drastically within a few megahertz. Nevertheless, 
relative intensities could be used frequently as guidance for assignments over several tens 
to a few hundreds of megahertz with only modest adjustments, in particular to check for 
possible blending.

\section{Spectroscopic properties of \textit{n}-propyl cyanide}
\label{rot_vib_backgr}

The molecule \textit{n}-propyl cyanide, also known as butyronitrile or 1-cyanopropane, exists 
in two distinguishable conformers, \textit{gauche} and \textit{anti}, with the former being 
doubly degenerate with respect to the latter. \citet{n-PrCN_IR_etc_2001} showed that 
\textit{gauche} is lower in energy than \textit{anti} by $40 \pm 3$~cm$^{-1}$.

The \textit{anti} conformer is an asymmetric rotor with $\kappa = (2B - A - C)/(A - C) = 
-0.9893$ close to the prolate limit of $-$1. Successive $R$-branch transitions are 
approximately spaced by $B + C \approx 4.4$~GHz. The $A$ rotational constant, $\sim$24~GHz, 
is a measure for the $K_a$ level spacing as the $K_a = J$ levels are roughly given by 
$A\,K_a^2$. The \textit{gauche} conformer is more asymmetric ($\kappa = -0.8471$) with 
a wider spacing between consecutive $R$-branch transitions ($B + C \approx 6.0$~GHz) 
and a narrower $K_a$ level spacing ($A \approx 10.1$~GHz).

The large dipole moment is caused predominantly by the CN group. \citet{n-PrCN_mmW_1988} 
determined the \textit{a-n}-PrCN dipole moment components as $\mu _a = 3.597 \pm 0.059$~D 
and $\mu _b = 0.984 \pm 0.015$~D; $\mu _c = 0$ by symmetry. \citet{i-PrCN_rot_2011} presented 
quantum chemical calculations on properties of \textit{i}-PrCN, \textit{a-n}-PrCN, and 
\textit{cyclo}-PrCN from which \citet{i-PrCN_det_2014} concluded that $\mu _a$ is likely 
underestimated by about 10\,\%. They proposed $\mu _a = 4.0$~D and suggested that the 
underestimation of $\mu _a$ could be responsible, at least in part, for the incorrect 
energy ordering of the conformers by \citet{n-PrCN_mmW_1988}. The results for 
\textit{g-n}-PrCN were $\mu _a = 3.272 \pm 0.037$~D and $\mu _b = 2.139 \pm 0.030$~D 
under the assumtion of $\mu _c = 0$ \citep{n-PrCN_mmW_1988}. Quantum chemical calculations 
yield $\mu _c \approx 0.45$~D \citep{n-PrCN_IR_etc_2001}, in line with the value 
mentioned by \citet{i-PrCN_det_2014}.

The orientation of the $^{14}$N quadrupole tensor in a CN group is usually very close 
to this bond. This leads to $\chi _{aa} = -3.44$~MHz and $\chi _{cc} = 2.06$~MHz for 
\textit{a-n}-PrCN and $\chi _{aa} = -1.68$~MHz and $\chi _{cc} = 1.94$~MHz for 
\textit{g-n}-PrCN \citep{n-PrCN_FTMW_1988}; $\chi _{bb}$ is determined from the 
tracelessness of the tensor.


\begin{figure} 
\centering 
\includegraphics[angle=0,width=8.8cm]{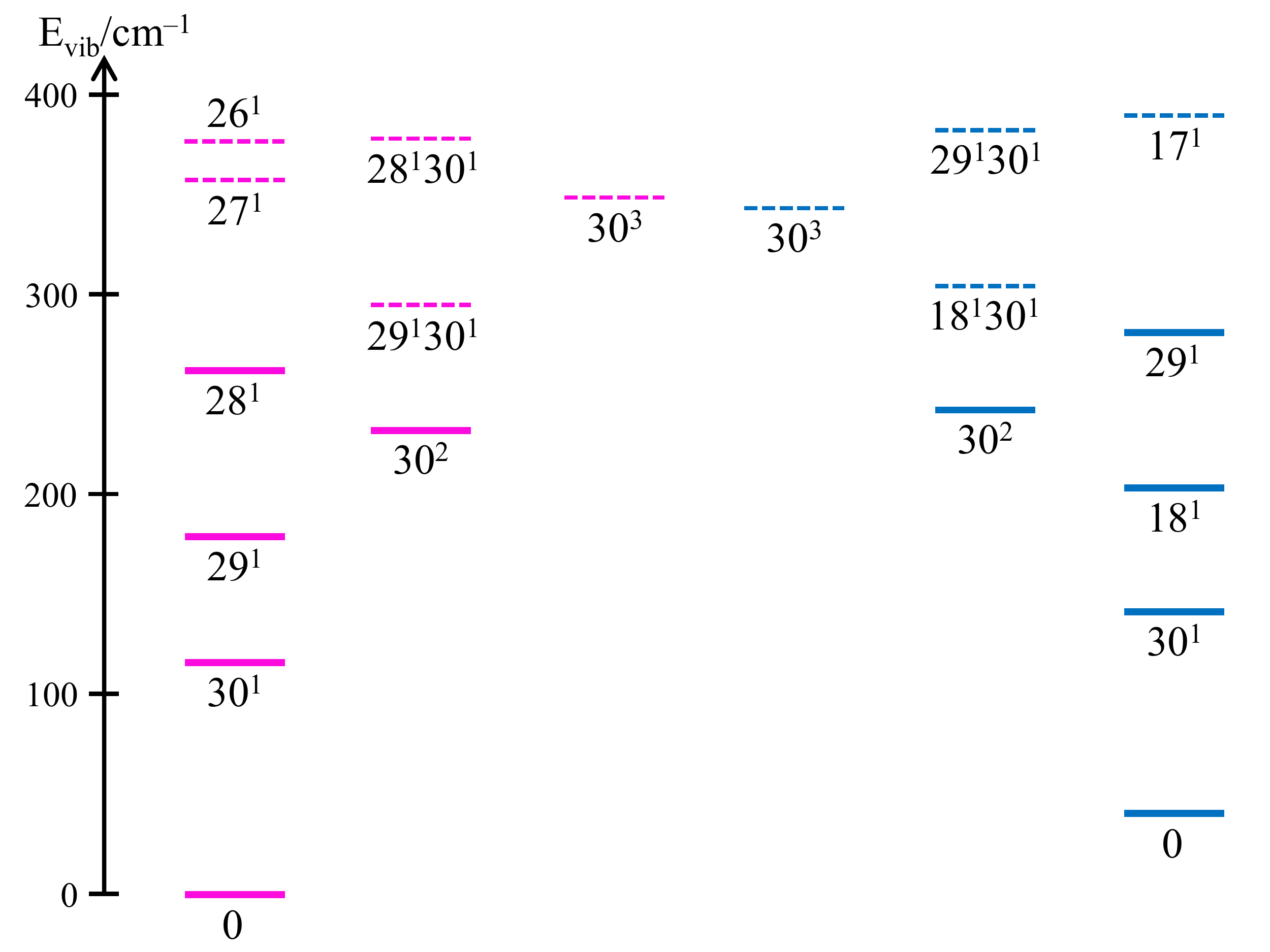} 
  \caption{Vibrational states of \textit{normal}-propyl cyanide up to vibrational energies 
           of 400~cm$^{-1}$. The \textit{gauche} states are shown on the left-hand side, those 
           of \textit{anti} on the right-hand side. The ground vibrational states and the four 
           vibrational states each of the present study are indicated by solid, thick lines, 
           additional states by dashed, thin lines.} 
\label{vibs} 
\end{figure} 


The \textit{n}-PrCN molecule has 12 atoms, its vibrational spectrum consists of 
$3 \times 12 - 6 = 30$ fundamental vibrations. The \textit{gauche} conformer has $C_1$ 
symmetry, so all vibrations belong to the symmetry class $a$. In contrast, the 
\textit{anti} conformer has $C_S$ symmetry, and 18 vibrations belong to the symmetry 
class $a'$ and 12 to the symmetry class $a''$ \citep{n-prCN_conf_IR_1987}. 
\citet{n-PrCN_IR_etc_2001} presented the most comprehensive study on the vibrational 
spectroscopy of this molecule. They measured the gas phase IR spectrum in the mid-IR 
(down to 300~cm$^{-1}$), and the solid-state IR and the liquid-state Raman spectra 
covered all fundamental vibrations. They supplemented these data by quantum chemical 
calculations at the MP2/6$-$31G* level; in addition, they presented scaled values 
for the fundamental vibrations to account for anharmonicity and model deficiencies. 
Figure~\ref{vibs} displays the vibrational states below 400~cm$^{-1}$.


\begin{table}
\begin{center}
\caption{Rotational partition function $Q_{\rm rot}$ and vibrational correction factors 
         $F_{\rm vib}$ to the rotational partition function values of \textit{n}-propyl 
         cyanide at selected temperatures $T$.}
\label{Q-values}
\begin{tabular}[t]{r@{}lr@{}lr@{}l}
\hline \hline
\multicolumn{2}{c}{$T$} & \multicolumn{2}{c}{$Q_{\rm rot}$}  & \multicolumn{2}{c}{$F_{\rm vib}$} \\
\multicolumn{2}{c}{(K)} & & \\
\hline
 300&.0   & 254102&.1111 & 9&.9477 \\
 225&.0   & 162159&.7064 & 4&.4642 \\
 180&.0   & 114139&.6161 & 2&.8253 \\
 150&.0   &  85494&.8293 & 2&.1160 \\
 120&.0   &  59864&.2360 & 1&.6192 \\
  75&.0   &  27985&.6804 & 1&.1684 \\
  37&.5   &   8997&.8511 & 1&.0129 \\
  18&.75  &   2968&.9170 & 1&.0001 \\
   9&.375 &   1032&.0718 & 1&.0000 \\
\hline
\end{tabular}
\end{center}
\end{table}


In order to evaluate contributions of vibrational states to the partition function $Q$ of 
\textit{n}-PrCN, we calculated vibrational correction factors $F_{\rm vib}$ to the rotational 
partition function values $Q_{\rm rot}$ ($Q = F_{\rm vib} \times Q_{\rm rot}$) at the 
seven CDMS standard temperatures plus two additional ones (120~K and 180~K). 
Scaled values from a quantum chemical calculation \citep{n-PrCN_IR_etc_2001} were taken 
for vibrational energies of the lowest three fundamental modes of the \textit{gauche} 
conformer and the gas phase experimental values from the same work for all others. 
Contributions of overtone and combination states were evaluated in the harmonic approximation. 
The resulting values of $Q_{\rm rot}$ and $F_{\rm vib}$ are given in Table~\ref{Q-values}.

Uncertainties in the vibrational energies, neglect of differences in the vibrational 
energies between the two conformers and the neglect of anharmonicity contribute 
to errors in $F_{\rm vib}$. We estimate these errors to be a few percent of 
$F_{\rm vib} - 1$ at low temperatures and possibly many tens of percent at 300~K.

A plethora of low-lying vibrational states of \textit{n}-PrCN are populated at 
room temperature such that only about one tenth of the molecules are in the 
ground vibrational states of the \textit{gauche} or \textit{anti} conformer. 
The lowest vibrational state ($\varv _{30} = 1$ for both conformers) has 
predominantly the character of a torsion around the central C atoms 
\citep{n-prCN_conf_IR_1976,n-prCN_conf_IR_1987,n-PrCN_IR_etc_2001} and could be 
called ethyl torsion. The next lowest state ($\varv _{29} = 1$ or $\varv _{18} = 1$ 
for \textit{gauche} and \textit{anti}, respectively) is a mixed bending vibration 
involving mostly the CCN angle and the CCC angle with the cyano-C. The next vibrational 
state is $\varv _{30} = 2$, followed by $\varv _{28} = 1$ (\textit{gauche}) or 
$\varv _{29} = 1$ (\textit{anti}) with predominantly methyl torsional character 
\citep{n-prCN_conf_IR_1976,n-prCN_conf_IR_1987,n-PrCN_IR_etc_2001}.

\section{Spectroscopic results and their discussion}
\label{lab-results}

We used Pickett's SPCAT and SPFIT programs \citep{spfit_1991} to predict and fit rotational 
spectra of vibrationally excited \textit{n}-PrCN. Vibrational changes $\Delta X$ of a 
spectroscopic parameter $X$ are usually considerably smaller in magnitude than the magnitude 
of $X$. Using ground state spectroscopic parameters and fitting the vibrational changes has 
two distinct advantages. First, the ground state parameters usually account for a substantial 
fraction of the excited state parameters. Second, the number of spectroscopic parameters used 
in the fit is considerably smaller than using independent sets of parameters for each state. 
Earlier examples include excited states of bromine dioxide \citep{OBrO_rot_1997}, chlorine 
dioxide \citep{OClO_rot_1997}, and the main and minor isotopologs of methyl cyanide 
\citep{MeCN_v8_le_2_rot_2015,MeCN_13C-vib_rot_2016}. We defined $\Delta X = X_{\varv = 1} - X_0$ 
and $\Delta \Delta X = X_{\varv = 2} - X_0 - 2\Delta X$. The ground state rotational and 
centrifugal distortion parameters of \textit{n}-PrCN were taken from \citet{det-PrCN_EtFo}, who 
combined data from earlier publications \citep{n-PrCN_rot_1982,n-PrCN_FTMW_1988,n-PrCN_mmW_1988},
and were kept fixed in the present analyses. Calculated values for $\Delta A$, $\Delta B$, and 
$\Delta C$ of each vibrational state of both conformers were derived from \citet{n-PrCN_rot_1962}.

\begin{figure*} 
\centering 
\includegraphics[angle=0,width=15.5cm]{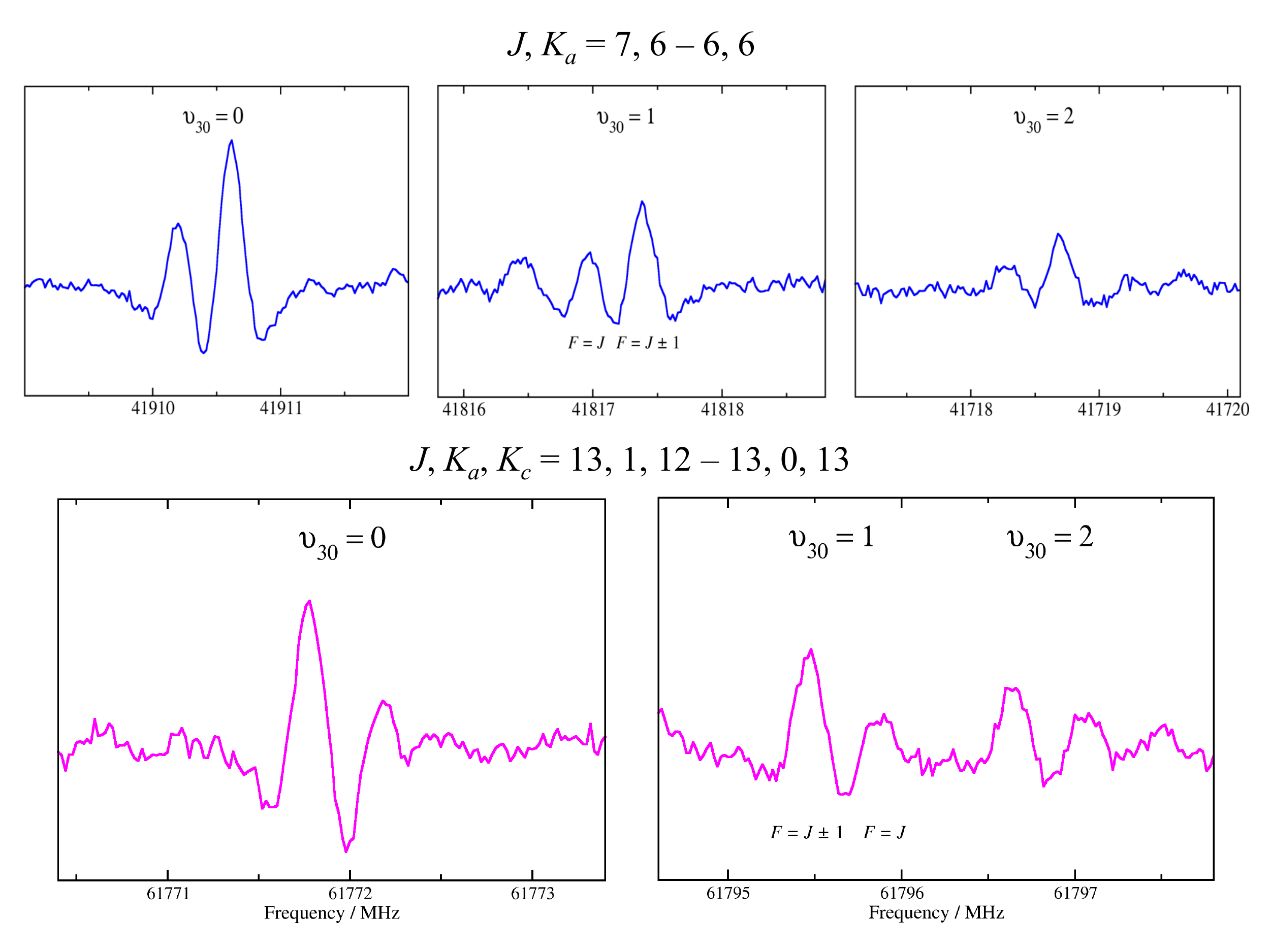} 
  \caption{Sections of the rotational spectrum of \textit{g-n}-PrCN showing transitions 
           with partially resolved $^{14}$N quadrupole splitting. The upper trace shows 
           the 7, 6 $-$ 6, 6 $a$-type $R$-branch transition with a 1:2 HFS pattern, 
           the lower trace shows the 13, 1, 12 $-$ 13, 0, 13 $b$-type $Q$-branch 
           transition with a 2:1 HFS pattern. Transitions for $\varv _{30} = 0$, 
           1, and 2 are shown from left to right in both cases; the vibrational labels 
           are given above the two HFS components, their assignments are indicated 
           below the $\varv _{30} = 1$ lines.}
\label{HFS} 
\end{figure*}

We recognized fairly early that a considerable fraction of the transitions in the 
laboratory spectrum, usually weaker ones, displayed partially resolved $^{14}$N 
hyperfine structure (HFS) splitting caused by the nuclear quadrupole moment of $^{14}$N, 
as is shown in Fig.~\ref{HFS}. Therefore, we generated predictions of the rotational 
spectra of excited states of \textit{n}-PrCN with and without considering HFS and 
compared the spectral recordings with the results of both predictions. The HFS parameters 
were taken from \citet{n-PrCN_FTMW_1988} and kept fixed in the fits.

The spin of the $^{14}$N nucleus ($I = 1$) leads to three HFS levels. Except for the 
lowest $J$ quanta, this leads to three stronger HFS components with $\Delta F = \Delta J$ and 
$F = J$, $J \pm 1$. The $F = J \pm 1$ components are almost always blended in our spectra such 
that a relatively well-resolved HFS pattern has an intensity ratio of nearly 2:1 
(see Fig.~\ref{HFS}). If the $F = J$ component is close to the $F = J \pm 1$ line, 
it appears weaker. This component was then given a larger uncertainty or it was not used 
in the fit if its position appeared to be affected too much by the proximity of the stronger 
$F = J \pm 1$ line. The $F = J \pm 1$ line could also appear weaker than expected 
if the two components were separated by several tens of kilohertz.


\subsection{The \textrm{gauche} conformer}
\label{gauche-results}

The \textrm{gauche} conformer has sizable $a$- and $b$-dipole moment components which 
facilitate assignments of corresponding transitions. Moreover, $a$-type transitions 
alone lead to fairly accurate values of $A$, $D_K$, etc. because the conformer is quite 
far away from the prolate symmetric limit. The asymmetry leads to a rapid increase of 
the asymmetry splitting for a given $K_a$ with increasing $J$, or, in other words, 
to resolved asymmetry splitting at relatively low values of $J$. In addition, transitions 
with $K_c = J''$ are almost degenerate (or oblate paired) close to 127~GHz.

We searched initially for $a$-type $R$-branch transitions ($\Delta J = +1$) with $K_a \le 2$ 
starting at low frequencies. These transitions were found easily close to the predicted 
frequencies. After refinement of the spectroscopic parameters, assignments were extended to 
higher $K_a$ values and to the 89$-$127~GHz frequency window. Once the assignments reached 
$K_a = 5$ we assigned $b$-type $R$-branch transitions with $K_a = 1 \leftrightarrow 0$ and 
then $K_a = 2 \leftrightarrow 1$. These transitions were mostly found easily and close to 
the predictions. Assignments of transitions pertaining to the $\varv _{28} = 1$ state 
were not as straightforward, as will be discussed in greater detail below.

The $a$-type $R$-branch transitions displayed partially resolved $^{14}$N HFS structure 
for the two highest values of $K_a$ ($J'' - 1$ and $J''$) in the 36$-$70~GHz window, 
as can be seen in the upper trace of Fig.~\ref{HFS}. The $b$-type transitions are also 
affected significantly by HFS splitting in certain quantum number ranges, especially 
$Q$-branch transitions ($\Delta J = 0$) with low values of $K_a$, see lower trace of 
Fig.~\ref{HFS}, as well as some $a$-type $Q$-branch transitions with $\Delta K_a = 0$ and 2.

\subsubsection{$\varv _{30} = 1$ and 2}
\label{g-v30_eq_1_2}

The quantum chemically calculated vibrational energy of $\varv _{30} = 1$ is 116~cm$^{-1}$ 
both scaled and unscaled \citep{n-PrCN_IR_etc_2001}. A shoulder in the Raman spectrum of 
the liquid sample at 114~cm$^{-1}$ agrees well with the calculated value, whereas the 
position of a strong feature in the solid-state IR spectrum at 140~cm$^{-1}$ agrees less 
well \citep{n-PrCN_IR_etc_2001}. The calculated energy of $\varv _{30} = 2$ in the harmonic 
approximations is 232~cm$^{-1}$. Anharmonicity is probably small; we will discuss this 
aspect in the context of $\varv _{30} = 2$ of the \textit{anti} conformer and its 
interaction with $\varv _{18} = 1$ in Sect.~\ref{a-reso}.

Figure~\ref{HFS} shows not only the partial HFS splitting, but also the decrease in 
intensity upon excitation of $\varv _{30}$. The $a$-type $R$-branch transitions reach 
$J = 23 - 22$ for $K_c = J$ near 127~GHz and $J = 21 - 20$ for $K_a \ge 3$. We assigned 
transitions up to $K_a = 20$ for both $\varv _{30} = 1$ and 2. $Q$-branch transitions 
with $\Delta K_a = 0$ or 2 are considerably weaker. Assignments with $\Delta K_a = 0$ 
reach $K_a = 11$ and $J = 61$ for $\varv _{30} = 1$ and $K_a = 8$ and $J = 45$ for 
$\varv _{30} = 2$. In the case of $\Delta K_a = 2$ we assign up to $K_a = 8 - 6$ and 
$J = 43$ for $\varv _{30} = 1$ and up to $K_a = 6 - 4$ and $J = 34$ for $\varv _{30} = 2$. 
The $b$-type transitions involve $^rR$, $^pR$, and $^rQ$ transitions up to $K_a = 11$ 
for both states and $J = 65$ and 63 for $\varv _{30} = 1$ and 2, respectively. 
The superscripts indicate $\Delta K_a = +1$ ($r$) and $-1$ ($p$), respectively. 
There was evidence for $b$-type transitions with even higher $K_a$ in the spectral 
recordings, but because of their weakness and greater sparseness we refrained from 
including them in the line lists at present. We point out that identification of 
$\varv _{30} = 2$ transitions was greatly facilitated, especially for very weak 
transitions, by $\varv _{30} = 1$ assignments already included in the fit because the 
ground state spectroscopic parameters and the $\varv _{30} = 1$ vibrational corrections 
(multiplied by two) accounted quite well for the $\varv _{30} = 2$ transition frequencies.

The vibrational changes of the parameters for $\varv _{30} = 1$ and the second changes 
for $\varv _{30} = 2$ are given in Table~\ref{gauche-parameters} together with ground state 
parameters and vibrational changes for $\varv _{29} = 1$ and $\varv _{28} = 1$. We determined 
a full set of vibrational changes up to fourth order along with several sixth order changes; 
the set of second changes is somewhat smaller, as can be expected. The quadrupole splitting 
showed systematic changes from $\varv _{30} = 0$ to 2 which permitted determination of a 
vibrational change to $\chi _{aa}$.

\subsubsection{$\varv _{29} = 1$}
\label{g-v29_eq_1}

\citet{n-PrCN_IR_etc_2001} give 163~cm$^{-1}$ as scaled calculated value, and the unscaled 
value is 1~cm$^{-1}$ higher. Their liquid- and solid-state Raman spectra display a medium 
strong band at 180~cm$^{-1}$, in moderate agreement. If we assume that results of quantum 
chemical calculations are similar in quality with respect to the values of the free, gaseous 
molecule of both conformers and that this also holds for the results from liquid- and 
solid-state Raman spectra, then our experimentally determined energy difference  between 
$\varv _{30} = 2$ and $\varv _{18} = 1$ (see Sect.~\ref{a-reso}) provides evidence that 
the vibrational energy of $\varv _{29} = 1$ of gaseous \textit{g-n}-PrCN is closer to 
163~cm$^{-1}$ than to 180~cm$^{-1}$.

The assignments of $a$-type $R$-branch transitions and of $b$-type transitions took place 
in a similar way as for $\varv _{30} = 1$ and 2. Assignment of weaker lines was often more 
difficult than in the case of $\varv _{30} = 2$ because larger uncertainties made unambiguous 
assignments more difficult even though corresponding $\varv _{30} = 2$ lines were slightly 
weaker; predictions of $\varv _{30} = 2$ frequencies were better because of the 
$\varv _{30} = 1$ assignments, see also Sect.~\ref{g-v30_eq_1_2}. Fewer $b$-type transitions 
extending not as high in $J$ are probably the most important reason why we have not been 
able to assign weak $a$-type $Q$-branch transitions in $\varv _{29} = 1$.

Excluding $\Delta \chi _{aa}$, the number of spectroscopic parameters in the fit of 
$\varv _{29} = 1$ is the same as that for $\varv _{30} = 1$, but the choice of sixth order 
parameters is different, see Table~\ref{gauche-parameters}.

\subsubsection{$\varv _{28} = 1$}
\label{g-v28_eq_1}

The CH$_3$ torsional mode is expected to be quite anharmonic. The unscaled vibrational energy 
for this mode is 276~cm$^{-1}$, and the scaled one 262~cm$^{-1}$ \citep{n-PrCN_IR_etc_2001}. 
These authors assigned a weak band at 265~cm$^{-1}$ in the IR spectrum of a solid sample, and 
also a very weak feature at 244~cm$^{-1}$ in the Raman spectrum of the liquid. We note, however, 
that the agreement of the latter vibrational energy agrees better with their scaled value for 
the corresponding mode of the \textit{anti} conformer, see also Sect.~\ref{a-v29_eq_1}.

Assignments of $a$-type $R$-branch transitions with $K_a \le 2$ were straightforward also for 
$\varv _{28} = 1$. Absorption features of higher $K_a$ transitions, however, often appeared 
somewhat weaker and broader than predicted and/or showed a weaker shoulder; some transitions 
even appeared split into two. We suspect that these features are caused by methyl internal 
rotation as $\varv _{28} = 1$ should possess considerable methyl torsion character 
\citep{n-prCN_conf_IR_1976,n-prCN_conf_IR_1987,n-PrCN_IR_etc_2001,n-PrCN_tors_2011}. 
The measured line center usually agreed sufficiently well with the calculated positions; 
however, in the case of a line that was split into two, the average position did not always 
agree well enough with the calculated position and was weighted out.

We were able to assign several $b$-type $R$-branch transitions with $K_a = 1 \leftrightarrow 0$ 
and $2 \leftrightarrow 1$. However, some of the lower $J$ transitions or $Q$-branch transitions 
with the same $K_a$ values were quite weak and appeared somewhat displaced from their predicted 
positions. Their frequencies could not be reproduced with one set of reasonable spectroscopic 
parameters. Assignment of $b$-type transitions with higher $K_a$ proved to be difficult; most 
promisingly, we made tentative assignments of several $K_a = 5 - 4$ $R$-branch transitions. 
However, it was difficult to fit these transitions together with the previous assignments. 
In addition, it was not possible to make any further unambiguous assignments.

The set of spectroscopic parameters for $\varv _{28} = 1$ is smaller than those for the other 
vibrational states of \textit{g-n}-PrCN. We estimated $\Delta D_K = - 2.2$~kHz to accomodate the 
fairly secure assignments of $b$-type transitions with $K_a \le 2$.

\subsection{The \textrm{anti} conformer}
\label{anti-results}

\begin{figure} 
\centering 
\includegraphics[angle=0,width=8.8cm]{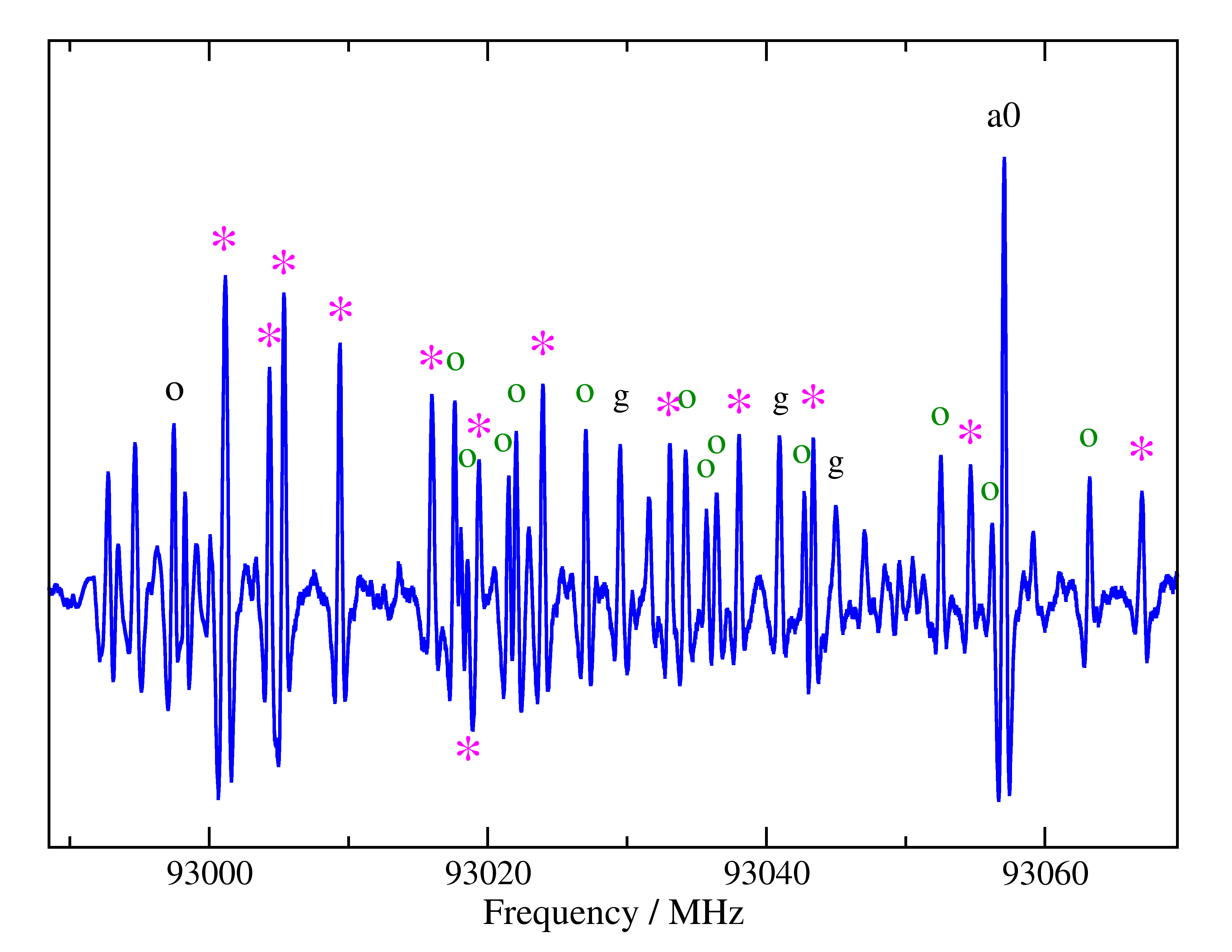} 
  \caption{Section of the rotational spectrum of \textit{n}-PrCN displaying 
           the clustering of $a$-type $R$-branch transitions of the \textit{anti} 
           conformer with higher values of $K_a$. Magenta asterisks indicate 
           transitions of $\varv _{30} = 1$, green circles those of $\varv _{18} = 1$, 
           a black circle a transition of $\varv _{30} = 2$, ''a0'' labels a 
           ground state transition, and a ''g'' labels transitions of the gauche 
           conformer.}
\label{anti} 
\end{figure}

The \textit{anti} conformer is much closer to the prolate symmetric top limit than 
\textit{gauche}, which makes the determination of $A$, $D_K$, etc. difficult from $a$-type 
transitions alone. The smaller magnitude of the $b$-dipole moment component of this conformer 
and its larger $A$ rotational constant make it more difficult to assign $b$-type transitions. 
Its much smaller asymmetry leads to unresolved asymmetry splitting for $K_a = 4$ at low 
frequencies and also throughout the two frequency windows for $K_a > 4$. Furthermore, most 
transitions ($K_a \ge 3$) of a given vibrational state occur in a narrow frequency range in 
the two frequency windows, as shown in Fig.~\ref{anti}, which often leads to blending of lines.

Initial assignments focussed on $K_a \le 2$ $a$-type $R$-branch transitions at lower 
frequencies, as in the case of the \textit{gauche} conformer. Assignments were extended 
to higher frequencies and to higher $K_a$ transitions with refined predictions. Overlap 
of transitions with different values of $K_a$ made the assignments less straightforward 
in the early stages of the assignment than for the \textit{gauche} conformer.

Splitting caused by the $^{14}$N quadrupole coupling affects a much larger fraction of 
the $a$-type $R$-branch transitions of the \textit{anti} conformer because the quadrupole 
coupling parameters are much larger in magnitude than for the \textit{gauche} conformer, 
most importantly $\chi _{aa}$. The $a$-type $Q$- or $P$-branch transitions of the 
\textit{anti} conformer were too weak to be identified.

Subsequently, we attempted to assign $b$-type transitions with $K_a = 1 \leftrightarrow 0$. 
These transitions were too weak in our spectral recordings covering 36 to 70~GHz, but 
sufficiently strong in the 89 to 127~GHz region. However, the density of lines at these 
low intensity levels together with the relatively large uncertainties made unambiguous 
assignments difficult. Therefore, we recorded smaller sections around the positions of 
the $b$-type transitions in the 36$-$70~GHz region with much longer integration time. 
In these recordings, we detected several candidate lines for each vibrational state. 
Further details will be provided below.

 
\subsubsection{$\varv _{30} = 1$}
\label{a-v30_eq_1}

The calculated scaled and unscaled vibrational energy of $\varv _{30} = 1$ is 101~cm$^{-1}$; 
this compares well with a strong band in the solid-state Raman spectrum at 99~cm$^{-1}$ 
\citep{n-PrCN_IR_etc_2001}.

The $a$-type transitions have $J = 9 - 8$ at low frequencies and $J = 28 - 27$ at high 
frequencies; in the case of $K_a = 1$, the $29_{1,29} - 28_{1,28}$ transition falls into 
the frequency ranges we have covered and analyzed thus far. The $K_a$ quantum numbers 
extend to 24, almost reaching the theoretical limit of 27 ($K_{a,{\rm max}} \le J_{\rm max}$). 
We were able to identify 10 $b$-type transitions in the smaller spectral recordings below 
70~GHz, the lowest frequency assignment was near 53~GHz. We assigned 14 additional $b$-type 
transitions between 89 and 127~GHz. The transitions are spread quite evenly among $^rR_0$, 
$^pR_1$, and $^rQ_0$ transitions; the subscript indicates $K_a$ in the lower energy rotational 
level, and the $J$ range covers 8 to 48.

We fit $\varv _{30} = 1$ and 2 together from the beginning. It required a modest set of 
vibrational changes to reproduce the $\varv _{30} = 1$ transitions well. These are changes 
to the rotational and quartic centrifugal distortion parameters and to $H_{KJ}$. 
We measured accurate frequencies of both $\Delta F = J \pm 1$ as well as $\Delta F = J$ 
for several transitions with higher $K_a$ of $\varv _{30} = 1$ and to a lesser extent for 
$\varv _{30} = 2$. In a trial fit, we obtained $\Delta \chi _{aa} = 10 \pm 35$~kHz which 
was omitted from the final fit. The resulting spectroscopic parameters are given in 
Table~\ref{anti-parameters} together with parameters of other vibrational states of the 
\textrm{anti} conformer.

Many of the vibrational changes $\Delta X$ are much smaller than the ground state parameters 
$X$ themselves, as is usually expected. The magnitude of $\Delta A$, however, is relatively 
large with respect to $\Delta B$ and $\Delta C$, even if we take into account that $A$ is 
about one order of magnitude larger than either $B$ or $C$. The $\Delta A$ value of 
$\varv _{18} = 1$ is of similar magnitude and of opposite sign, suggesting the large 
magnitudes are caused to a considerable amount by the non-resonant Coriolis interaction 
between these states. Strong, albeit resonant, Coriolis interaction was identified between 
the corresponding modes $\varv _{13} = 1$ and $\varv _{21} = 1$ of the lighter homolog 
ethyl cyanide \citep{EtCN_rot_1959}. We have carried out trial fits that have shown that 
$|\zeta _a| \approx 0.82$ yields $\Delta A \approx -37$~MHz for both states and affects 
several of the vibrational changes of the distortion parameters considerably. 
Accurate harmonic force field calculations would be required to establish the actual 
value of $|\zeta _a|$ as well as the non-zero value of $|\zeta _b|$. Such calculations 
are beyond the aim of the current study.

\subsubsection{$\varv _{29} = 1$}
\label{a-v29_eq_1}

\citet{n-PrCN_IR_etc_2001} calculated an unscaled value of 254~cm$^{-1}$ for the vibrational 
energy of the CH$_3$ torsional mode and a scaled value of 241~cm$^{-1}$. They do not assign 
any experimental feature to this mode. We point out that a very weak band in the Raman 
spectrum of the liquid sample at 244~cm$^{-1}$ agrees well with the scaled value; 
\citet{n-PrCN_IR_etc_2001} assigned this band to the CH$_3$ torsional mode of the 
\textrm{gauche} conformer, see also Sect.~\ref{g-v28_eq_1}.

The assignments of $a$-type transitions pertaining to $\varv _{29} = 1$ proceeded analogously 
to those of $\varv _{30} = 1$, but extended only to $K_a = 18$ mainly because it is the 
highest of the vibrational states of \textit{n}-PrCN in the present study. As can be seen 
in Fig.~\ref{anti}, not only do high-$K_a$ transitions of each vibrational state of 
\textit{a-n}-PrCN occur in a narrow frequency region, but also those of other vibrational 
states are often very close. Therefore, blending of lines reduced the number of assignments. 
For example only $J = 28 - 27$ could be seen as apparently not blended in the case of 
$K_a = 17$. Even though there are two vibrational states of \textit{a-n}-PrCN close to 
this state, $\varv _{18} = \varv _{30} = 1$ and $\varv _{30} = 2$, we did not find clear 
evidence for perturbations of $\varv _{29} = 1$.

We found six lines with approximately the right intensity close to the predicted frequencies 
of $K_a = 1 \leftrightarrow 0$ $b$-type transitions in the long integration, narrow spectral 
recordings between 54 and 67~GHz. These lines could be fit within experimental uncertainties 
on average. With improved predictions, six more transitions close to the predicted positions 
were tentatively assigned in the 89 to 127~GHz region. The twelve $b$-type transitions, however, 
could not be fit within experimental uncertainties with a reasonable set of spectroscopic 
parameters. It is possible that some positions are affected by unrecognized blending. Even though 
$\varv _{29} = 1$ should have predominantly methyl torsion character, we did not find any clear 
evidence for broadening or even splitting of lines due to torsion. Nevertheless, we do not rule 
out that the unsatisfactory fitting of the $b$-type transitions may be related to torsional 
effects. We have omitted the lines from the final line list because of this unclear situation.

The set of spectroscopic parameters in Table~\ref{anti-parameters} is much smaller for 
$\varv _{29} = 1$ than it is for $\varv _{30} = 1$; only changes for the rotational parameters, 
$D_{JK}$, $D_J$, and $H_{KJ}$ were needed to fit the $a$-type transitions of $\varv _{29} = 1$ 
satisfactorily.

\subsubsection{$\varv _{18} = 1$ and $\varv _{30} = 2$}
\label{a-reso}

Whereas \citet{n-PrCN_IR_etc_2001} measured a medium strong band in the Raman spectrum of liquid 
and solid samples at 180~cm$^{-1}$, they report a scaled calculated value of 163~cm$^{-1}$; the 
unscaled value is 1~cm$^{-1}$ higher. The calculated vibrational energy of $\varv _{30} = 2$ is 
202~cm$^{-1}$ in the harmonic approximation.

 
\begin{figure}
\centering
\includegraphics[angle=0,width=5.4cm]{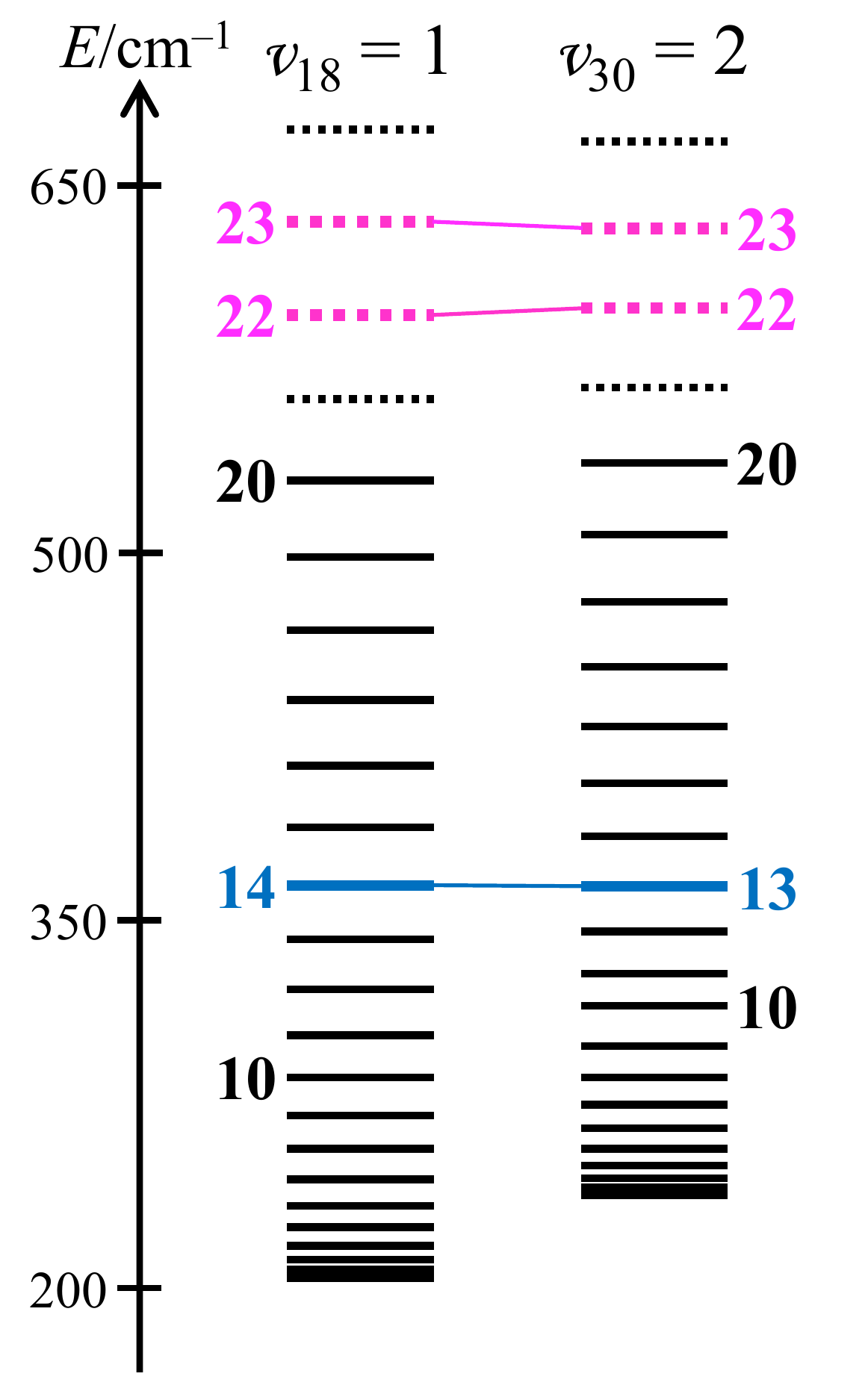}
  \caption{$K_a$ level structure of $\varv _{18} = 1$ and $\varv _{30} = 2$ 
           of \textit{a-n}-PrCN displaying the $\Delta K_a = 1$ rotational 
           (Coriolis-type) resonance as well as the $\Delta K_a = 0$ Fermi resonance. 
           The most strongly interacting levels are shown with colored lines and are 
           connected by thin lines. Dashed lines indicate $K_a$ levels not accessed 
           in the present study.}
\label{resonance}
\end{figure}


\begin{table*}
\begin{center}
\caption{Spectroscopic parameters $X^a$ (MHz) of \textit{gauche-n}-propyl cyanide and first ($\Delta X$) and second 
         ($\Delta \Delta X$) changes$^b$ of low-lying vibrational states.}
\label{gauche-parameters}
\renewcommand{\arraystretch}{1.10}
\begin{tabular}[t]{lr@{}lr@{}lr@{}lr@{}lr@{}l}
\hline \hline
Parameter & \multicolumn{2}{c}{$\varv = 0$} & \multicolumn{2}{c}{$\varv_{30} = 1$} & \multicolumn{2}{c}{$\varv_{30} = 2$} 
  & \multicolumn{2}{c}{$\varv_{29} = 1$} & \multicolumn{2}{c}{$\varv_{28} = 1$} \\
          & \multicolumn{2}{c}{$X$} & \multicolumn{2}{c}{$\Delta X$} & \multicolumn{2}{c}{$\Delta \Delta X$} 
  & \multicolumn{2}{c}{$\Delta X$} & \multicolumn{2}{c}{$\Delta X$} \\
\hline
$A$                      &   10~060&.416~5(11)   &   58&.641~9(3)  &    3&.732~3(5)   &    23&.723~7(5)   & $-$36&.927~9(135) \\
$B$                      &    3~267&.662~41(30)  & $-$7&.938~29(5) & $-$0&.603~51(11) &     2&.493~94(13) &     2&.252~06(85) \\
$C$                      &    2~705&.459~57(29)  & $-$5&.173~90(5) & $-$0&.159~73(11) &     0&.431~86(12) &  $-$0&.175~62(44) \\
$D_K \times 10^3$        &       60&.235(6)      &    5&.005(4)    &    0&.782(5)     &  $-$0&.829(4)     &   $-$2&.2$^c$     \\
$D_{JK} \times 10^3$     &    $-$18&.264~7(12)   & $-$0&.656~7(3)  & $-$0&.138~5(3)   &     0&.155~7(8)   &      0&.335~9(23) \\
$D_J \times 10^6$        &    3~195&.06(21)      &    7&.76(4)     &    6&.36(8)      & $-$20&.64(22)     &      8&.94(57)    \\
$d_1 \times 10^6$        & $-$1~037&.470(55)     & $-$8&.029(10)   & $-$2&.455(22)    &     1&.804(33)    &   $-$5&.61(49)    \\
$d_2 \times 10^6$        &    $-$77&.186(18)     & $-$3&.605(3)    & $-$0&.280(7)     &  $-$1&.244(18)    &   $-$1&.19(27)    \\
$H_K \times 10^6$        &        1&.806(18)     &    0&.476(21)   &     &            &      &            &       &           \\
$H_{KJ} \times 10^6$     &     $-$0&.517~3(35)   & $-$0&.109~5(10) &     &            &  $-$0&.042~8(41)  &       &           \\
$H_{JK} \times 10^9$     &        9&.92(68)      &    5&.52(8)     &     &            &  $-$6&.26(27)     &       &           \\
$H_J \times 10^9$        &        4&.486(56)     &     &           &     &            &  $-$0&.524(95)    &       &           \\
$h_1 \times 10^9$        &        2&.505(29)     & $-$0&.148(3)    & $-$0&.159(5)     &      &            &       &           \\
$h_2 \times 10^{12}$     &      525&.(14)        &     &           &     &            &   180&.(15)       &       &           \\
$h_3 \times 10^{12}$     &      111&.3(31)       &    5&.0(5)      &     &            &    48&.1(64)      &       &           \\
$L_{KKJ} \times 10^{12}$ &       30&.6(31)       &     &           &     &            &      &            &       &           \\
$L_{JK} \times 10^{12}$  &     $-$4&.11(78)      &     &           &     &            &      &            &       &           \\
$\chi _{aa}$             &     $-$1&.683(4)      & $-$0&.061(10)   &     &            &      &            &       &           \\
$\chi _{bb}$             &     $-$0&.252(5)$^d$  &     &           &     &            &      &            &       &           \\
$\chi _{cc}$             &        1&.935(5)$^d$  &     &           &     &            &      &            &       &           \\
\hline
\end{tabular}\\[2pt]
\end{center}
\tablefoot{
$^{(a)}$ Watson's $S$ reduction has been used in the representation $I^r$. Ground state parameters are 
         from \citet{det-PrCN_EtFo} except quadrupole parameters which are from \citet{n-PrCN_FTMW_1988}. 
         Numbers in parentheses are one standard deviation in units of the least significant figures. 
         Parameters without uncertainties were estimated and kept fixed in the analyses. 
$^{(b)}$ $\Delta X = X_{\varv = 1} - X_0$; $\Delta \Delta X = X_{\varv = 2} - X_0 - 2\Delta X$. 
$^{(c)}$ Estimated value, see Sect.~\ref{g-v28_eq_1}. 
$^{(d)}$ Derived values. 
}
\end{table*}


\begin{table*}
\begin{center}
\caption{Spectroscopic parameters $X^a$ (MHz) of \textit{anti-n}-propyl cyanide and first ($\Delta X$) and second 
         ($\Delta \Delta X$) changes$^b$ of low-lying vibrational states and parameters describing the interaction 
         between $\varv_{18} = 1$ and $\varv_{30} = 2$.}
\label{anti-parameters}
\renewcommand{\arraystretch}{1.10}
\begin{tabular}[t]{lr@{}lr@{}lr@{}lr@{}lr@{}l}
\hline \hline
Parameter & \multicolumn{2}{c}{$\varv = 0$} & \multicolumn{2}{c}{$\varv_{30} = 1$} & \multicolumn{2}{c}{$\varv_{30} = 2$} 
  & \multicolumn{2}{c}{$\varv_{18} = 1$} & \multicolumn{2}{c}{$\varv_{29} = 1$} \\
          & \multicolumn{2}{c}{$X$} & \multicolumn{2}{c}{$\Delta X$} & \multicolumn{2}{c}{$\Delta \Delta X$} 
  & \multicolumn{2}{c}{$\Delta X$} & \multicolumn{2}{c}{$\Delta X$} \\
\hline
$A$                           &   23~668&.319~3(14)   & $-$810&.855(12)   &   93&.495(77)    &   738&.035(75)    &  $-$0&.702(137)   \\
$B$                           &    2~268&.146~89(15)  &      1&.451~97(7) &    0&.163~73(12) &     5&.158~11(20) &  $-$2&.318~97(14) \\
$C$                           &    2~152&.963~95(17)  &      5&.005~71(6) &    0&.158~67(29) &     1&.975~84(33) &  $-$1&.495~34(14) \\
$D_K \times 10^3$             &      240&.65(3)       & $-$263&.6(80)     &     &            &      &            &       &           \\
$D_{JK} \times 10^3$          &    $-$10&.826~3(9)    &      0&.496~8(5)  &    0&.138~0(12)  &  $-$0&.354~6(11)  &      0&.092~4(9)  \\
$D_J \times 10^6$             &      398&.67(7)       &      7&.94(3)     &     &            &  $-$1&.80(8)      &   $-$0&.45(6)     \\
$d_1 \times 10^6$             &    $-$46&.64(4)       &      1&.45(3)     &     &            &  $-$1&.14(9)      &       &           \\
$d_2 \times 10^6$             &     $-$0&.590(6)      &      2&.416(24)   &     &            &  $-$1&.760(52)    &       &           \\
$H_K \times 10^6$             &        2&.5           &       &           &     &            &      &            &       &           \\
$H_{KJ} \times 10^9$          &      372&.4(24)       &     63&.5(14)     &   17&.$^c$       &      &            &     63&.1(39)     \\
$H_{JK} \times 10^9$          &    $-$20&.67(20)      &       &           &     &            &      &            &       &           \\
$H_J \times 10^9$             &        0&.353(11)     &       &           &     &            &      &            &       &           \\
$h_1 \times 10^9$             &        0&.117(14)     &       &           &     &            &      &            &       &           \\
$\chi _{aa}$                  &     $-$3&.440(4)      &       &           &     &            &      &            &       &           \\
$\chi _{bb}$                  &        1&.385(5)$^d$  &       &           &     &            &      &            &       &           \\
$\chi _{cc}$                  &        2&.055(5)$^d$  &       &           &     &            &      &            &       &           \\
Interaction $\varv_{18} = 1$/$\varv_{30} = 2$ & &     &       &           &     &            &      &            &       &           \\
$E(30^2 - 18) \times 10^{-3}$ &         &             &       &           & 1017&.98(61)     &      &            &       &           \\
$F(18,30^2) \times 10^{-3}$   &         &             &       &           &   36&.48(44)     &      &            &       &           \\
$G_c(18,30^2)$                &         &             &       &           &  129&.32(16)     &      &            &       &           \\
\hline
\end{tabular}\\[2pt]
\end{center}
\tablefoot{
$^{(a)}$ Watson's $S$ reduction has been used in the representation $I^r$. Ground state parameters are 
         from \citet{det-PrCN_EtFo} except quadrupole parameters which are from \citet{n-PrCN_FTMW_1988}. 
         Numbers in parentheses are one standard deviation in units of the least significant figures. 
         Parameters without uncertainties were estimated and kept fixed in the analyses. 
$^{(b)}$ $\Delta X = X_{\varv = 1} - X_0$; $\Delta \Delta X = X_{\varv = 2} - X_0 - 2\Delta X$. 
$^{(c)}$ Estimated value, see Sect.~\ref{a-reso}. 
$^{(d)}$ Derived values. 
}
\end{table*}


The initial assignments of $\varv _{30} = 2$ and $\varv _{18} = 1$ $a$-type 
transitions proceeded in a similar way to those of $\varv _{30} = 1$ and 
$\varv _{29} = 1$. However, fitting transitions of $\varv _{30} = 2$ with 
$K_a = 10$ and 11 required a value of $\Delta \Delta H_{KJ}$ that was 
larger than $\Delta H_{KJ}$ of $\varv _{30} = 1$. Furthermore, this 
parameter was not sufficient to reproduce transitions having $K_a = 12$ 
and 13, and transitions with even higher $K_a$ could not be assigned 
with confidence. Weighting out all lines with $K_a \ge 10$ and constraining 
$\Delta \Delta H_{KJ}$ such that its ratio with $\Delta H_{KJ}$ is 
equal to the $\Delta \Delta D_{JK}$ to $\Delta D_{JK}$ ratio showed 
that the $K_a = 13$ transitions had residuals between 1.56 and 2.95~MHz 
for $20 \le J'' \le 27$. These residuals were almost perfectly matched 
by residuals between $-$1.65 and $-$2.95~MHz in $K_a = 14$ of $\varv _{18} = 1$, 
indicative of a $\Delta K_a = \pm 1$ rotational (also known as Coriolis-type) 
resonance between these states. Figure~\ref{resonance} displays the $K_a$ 
level structure of these states with the resonant interactions highlighted. 
We estimated the energy difference $E(30^2 - 18)$ between the two states and 
adjusted the first order Coriolis parameter $G_c$. After both parameters were 
fit and updated, predictions were generated, and we were able to assign transitions 
with higher $K_a$ for both states. However, they soon showed deviations rapidly 
increasing with $K_a$ in opposite directions for the two vibrations, indicative 
of a Fermi resonance. After inclusion of a Fermi parameter $F$ in the fit, 
assignments could be extended up to $K_a = 20$. Transitions with even higher 
$K_a$ were too weak to be assigned with confidence.

We found ten $K_a = 1 \leftrightarrow 0$ $b$-type transitions in the long 
integration, narrow spectral recordings between 38 and 70~GHz for 
$\varv _{18} = 1$ and seven between 54 and 67~GHz for $\varv _{30} = 2$; 
nine and seven more, respectively, were found between 89 and 127~GHz. 
The $J$ range covers 13$-$30 and 8$-$31 for $\varv _{18} = 1$ and 
$\varv _{30} = 2$, respectively; two tentative assignments with $J = 47$ 
and 48 for $\varv _{30} = 2$ were weighted out in the final line list.

The set of vibrational changes to the ground state spectroscopic parameters is 
slightly smaller for $\varv _{18} = 1$ than it is for $\varv _{30} = 1$. 
In the case of $\varv _{30} = 2$, second changes were used for the rotational 
parameters and for $D_{JK}$; in addition, an estimate of $\Delta \Delta H_{KJ}$ 
was used as fixed parameter as described above.

The experimentally determined energy difference of $33.956 \pm 0.020$~cm$^{-1}$ 
between these two states is quite well-matched by the calculated difference of 
39~cm$^{-1}$, but less so by 18~cm$^{-1}$ ($2 \times 99 - 180$) derived from 
solid-state Raman measurements \citep{n-PrCN_IR_etc_2001}. If the vibrational 
energy of $\varv _{30} = 2$ is close to 202~cm$^{-1}$, that of $\varv _{18} = 1$ 
is equally close to 168~cm$^{-1}$. \citet{n-PrCN_tors_2011} studied the 
anharmonicity of the methyl and ethyl torsional modes. The harmonic and 
anharmonic energies of the methyl torsions differ by more than 15\% for 
both conformers. The ethyl torsion was calculated to be marginally 
anharmonic in all models in case of the \textit{gauche} conformer, whereas 
the results are less clear for the \textit{anti} conformer. It is only slightly 
anharmonic in one model, but displays pronounced negative anharmonicity 
(i.e., the energy of $\varv _{30} = 2$ is \textit{higher} than twice the energy 
of $\varv _{30} = 1$) in two others. The vibrational energy of $\varv _{30} = 2$ 
in all models is around 225~cm$^{-1}$, which appears to be too high and higher 
than the calculated anharmonic energy of 209.54~cm$^{-1}$ for $\varv _{29} = 1$, 
which appears to be too low. The difference between the anharmonic energies of 
$\varv _{30} = 2$ and the (harmonic) energy of $\varv _{18} = 1$ from 
\citet{n-PrCN_tors_2011} yields values between $\sim$63 and $\sim$68~cm$^{-1}$, 
much larger than our experimental value of $\sim$34~cm$^{-1}$. Gas phase 
far-infrared measurements may be the most promising way to establish the energies 
of the lowest vibrational states of \textit{n}-PrCN.


\section{Results of astronomical observations}
\label{astro-results}

Part of the observations used in this article were briefly described in 
\citet{i-PrCN_det_2014}. A detailed account of the observations, reduction, 
and analysis method of the full data set covering 84.1 to 114.4~GHz was 
reported by \citet{deuterated_SgrB2N2_2016}. Here we use this full data set.

As already reported in \citet{i-PrCN_det_2014} on the basis of a smaller set 
of data, \textit{n}-propyl cyanide is clearly detected in its vibrational 
ground state toward Sgr~B2(N2), with about 120 lines detected with little 
contamination in the full EMoCA survey (Table~\ref{t:ndet} and 
Fig.~\ref{f:spec_c3h7cn-n_ve0}). A source size of 1.0$\arcsec$ was derived by 
\citet{i-PrCN_det_2014} from Gaussian fits to the integrated intensity maps of 
all detected lines that suffered the least from contamination by other 
species. The spectrum was well-fit under the local thermodynamic 
equilibrium (LTE) approximation with a temperature of 150~K, a linewidth of 
5~km~s$^{-1}$, and a velocity of 73.5~km~s$^{-1}$, leading to a column density 
of $1.7 \times 10^{17}$~cm$^{-2}$ after correction for the contribution of 
vibrationally excited states to the partition function. Please note that the 
column density is slightly different compared to the one reported by 
\citet{i-PrCN_det_2014}, $1.8 \times 10^{17}$ cm$^{-2}$, because we use here 
a slightly different vibrational correction factor to the rotational 
partition function (see Table~\ref{Q-values}). The lines of the vibrational 
ground state of \textit{n}-propyl cyanide detected above 110.7~GHz that were 
not reported in \citet{i-PrCN_det_2014} are well-fit with these parameters 
too (see last two pages of Fig.~\ref{f:spec_c3h7cn-n_ve0}).

Using the same LTE parameters as for the vibrational ground state, we looked 
for lines from within vibrationally excited states of both \textit{gauche} and 
\textit{anti} propyl cyanide in the EMoCA survey of Sgr~B2(N2). On the basis 
of this LTE model, we report the detection of one or several transitions or 
groups of transitions of each of the eight vibrationally excited states 
analyzed in the previous sections (Table~\ref{t:ndet}). The parameters of 
the detected lines are listed in 
Tables~\ref{t:list_g_v30e1}--\ref{t:list_a_v29e1}, and the spectra of these 
lines plus all those that contribute significantly to the signal detected 
toward Sgr~B2(N2) are shown in 
Figs.~\ref{f:spec_c3h7cn-n-g_v30e1}--\ref{f:spec_c3h7cn-n-a_v29e1}. 
The fact that the lines of all states, including the vibrational ground state, 
are well-fit with the same parameters, gives us confidence in the reliability 
of the identifications, even for the states with only one clearly detected line. 
The median size of all detected ground state and vibrationally excited state 
lines that do not suffer from contamination remains 1.0$\arcsec$, with 
an rms dispersion of $\sim$0.5$\arcsec$ and no obvious correlation with energy 
level.

\begin{figure*}
\centerline{\resizebox{0.7\hsize}{!}{\includegraphics[angle=0]{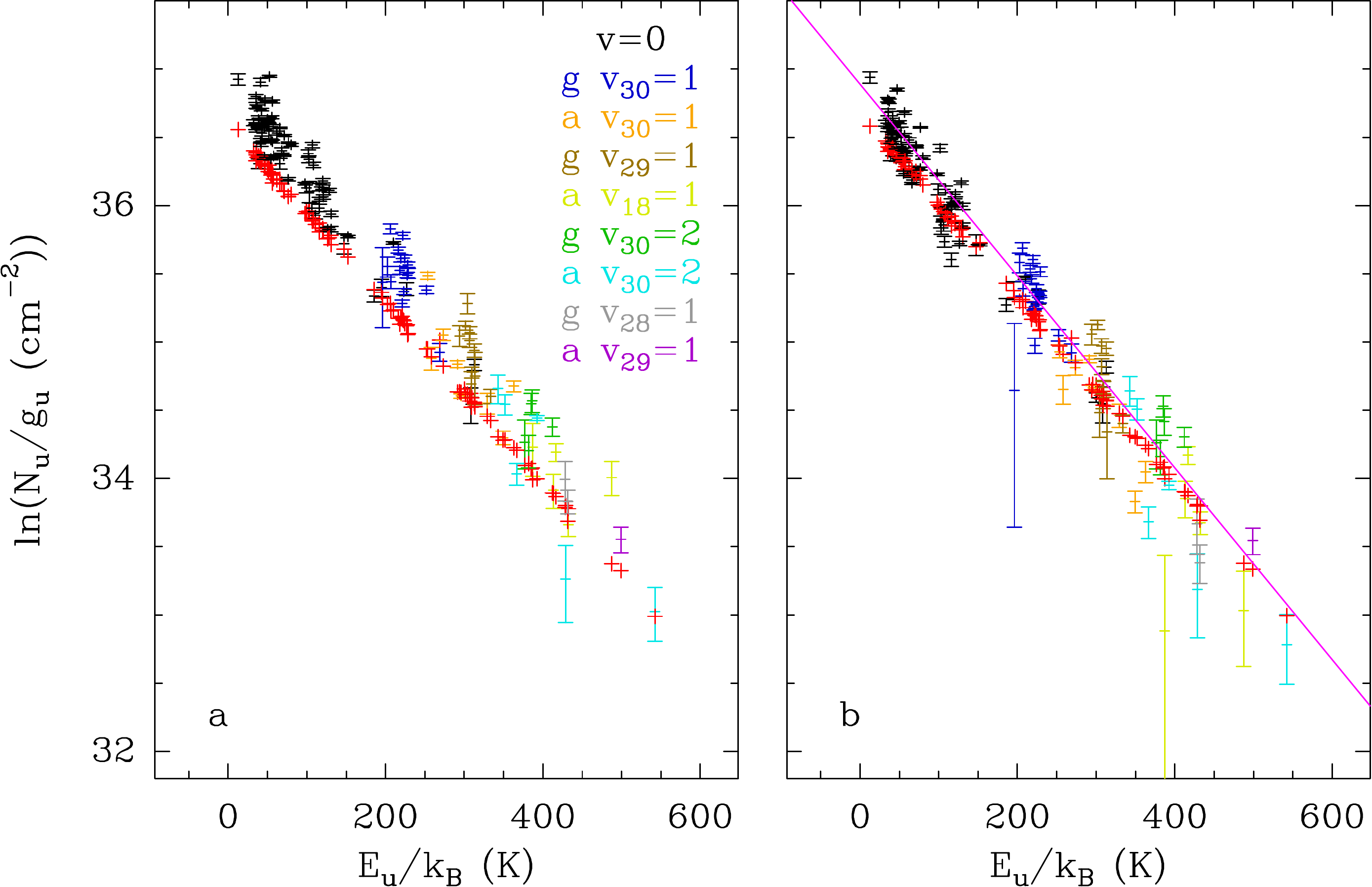}}}
\caption{Population diagram of \textit{n}-propyl cyanide toward Sgr~B2(N2). 
Only the lines that are clearly detected and do not suffer too much from 
contamination by other species are displayed. The observed data points are 
shown in various colors (except red) as indicated in the upper right corner 
of panel \textbf{a} while the synthetic populations are shown in red. No 
correction is applied in panel \textbf{a}. 
In panel \textbf{b}, the optical depth correction has been applied to both the 
observed and synthetic populations and the contamination by all other 
species included in the full model has been removed from the observed 
data points. The purple line is a linear fit to the observed populations (in 
linear-logarithmic space). The derived rotation temperature is $142 \pm 4$~K.
}
\label{f:popdiag}
\end{figure*}

The population diagram of most detected lines plus those that are somewhat more
contaminated but for which we have already identified and modeled most of the 
contaminants is shown in Fig.~\ref{f:popdiag}. The detected lines that consist 
of a blend of several transitions with significantly different upper-level 
energies are not used in this diagram. A fit to this diagram yields a
temperature of 142 $\pm$ 4~K. As explained in Sect.~3 of 
\citet{deuterated_SgrB2N2_2016}, the uncertainty on the fitted rotation
temperature is only statistical and does not include the systematic 
uncertainties due in particular to residual contamination by unidentified 
species. Our LTE model of \textit{n}-propyl cyanide assumes a temperature of
150~K, which is consistent with the fitted rotation temperature within 
2$\sigma$. The LTE model is optimized so that it never overestimates the peak 
temperature of any detected transition of \textit{n}-propyl cyanide, therefore 
it tends to produce synthetic integrated intensities that lie below the 
measured ones which may suffer from residual contamination, in particular in 
the wings of the lines (see the location of the red crosses with respect to 
the other ones in Fig.~\ref{f:popdiag}). This is the reason why our LTE model 
does not use the temperature strictly derived from the population diagram.

\input{tab_c3h7cn-n_ndet.tex}

\section{Conclusion and outlook}
\label{conclusion}

We have analyzed the rotational spectra of the four lowest excited vibrational states of 
each of the two \textit{n}-PrCN conformers in laboratory spectral recordings up to 127~GHz. 
We identified rovibrational interactions between $\varv _{18} = 1$ and $\varv _{30} = 2$ 
of the \textit{anti} conformer. We modeled the perturbations with one Coriolis and one 
Fermi parameter and determined the energy difference between these states quite accurately.

The resulting spectroscopic parameters enabled us to identify transitions of each excited 
vibrational state in our ALMA 3~mm molecular line survey of Sgr~B2(N2). The emission of all 
states is well reproduced with the model parameters we obtained previously for the ground 
vibrational state.

The emission caused by molecules in excited vibrational states may be used to infer 
(far-) infrared pumping in a given source. Our results concerning vibrational states of 
\textit{n}-PrCN suggest that far-infrared pumping is, at least, not so important that 
it alters the apparent vibrational temperature in the region of Sgr~B2(N2) which is 
probed by the \textit{n}-PrCN emission. We note that in Sgr~B2(N2) there are molecules 
which are better suited to investigate (far-) infrared pumping, even more so in other 
sources in which \textit{n}-PrCN has not yet been detected.

If LTE is a reasonable assumption, transitions of molecules in excited vibrational states 
may be used to constrain the rotational temperature if ground state rotational transitions 
cover an energy range that is too narrow to constrain it. This may be useful for linear 
molecules or for fairly light molecules; for example, we used transitions of torsionally 
excited methanethiol to constrain the rotational temperature of this molecule 
\citep{RSH_ROH_SgrB2N2_2016}.

Observations with deeper integration may make it possible to identify even higher 
vibrationally excited states of \textit{n}-PrCN as long as the confusion limit is 
not reached. In addition, the sensitivity of our current ALMA data may be sufficient 
to identify the $^{13}$C isotopomers of this molecule because the $^{12}$C/$^{13}$C ratio 
of many molecules in Sgr~B2(N2) was determined to be close to 25 
\citep{deuterated_SgrB2N2_2016,RSH_ROH_SgrB2N2_2016,EtCN_2x13C_2016}.

Predictions of the rotational spectra of the four lowest excited vibrational 
states of both conformers of \textit{n}-PrCN will be available in the catalog 
section\footnote{website: http://www.astro.uni-koeln.de/cdms/entries/}  
of the ascii version of the Cologne Database for Molecular Spectroscopy (CDMS) 
\citep{CDMS_1,CDMS_2} as well as in the Virtual Atomic and Molecular Data 
Centre (VAMDC) \citep{VAMDC_2010,VAMDC_2016} compatible version of the 
CDMS\footnote{http://cdms.ph1.uni-koeln.de/cdms/portal} 
\citep{CDMS_3}. The complete line, parameter, and fit files along with auxiliary 
files will be deposited in the Spectroscopy Data section of the 
CDMS\footnote{http://www.astro.uni-koeln.de/site/vorhersagen/daten/n-PrCN/}.

Future laboratory work will focus on analyzing the submillimeter spectra to extend 
assignments for the eight vibrationally excited states of the present investigation. 
This work has begun, should improve spectroscopic parameters considerably, and will 
be useful for potential assignments of radio astronomical spectra at higher frequencies. 
Furthermore, we will try to assign higher excited vibrational states; initial assignments 
exist for some combination and overtone states. Such states may be observable in more 
sensitive ALMA data of Sgr~B2(N).


\begin{acknowledgements}
This paper makes use of the following ALMA data: 
ADS/JAO.ALMA\#2011.0.00017.S, ADS/JAO.ALMA\#2012.1.00012.S. 
ALMA is a partnership of ESO (representing its member states), NSF (USA) and NINS (Japan), 
together with NRC (Canada), NSC and ASIAA (Taiwan), and KASI (Republic of Korea), 
in cooperation with the Republic of Chile. The Joint ALMA Observatory is operated by ESO, 
AUI/NRAO and NAOJ. The interferometric data are available in the ALMA archive at
https://almascience.eso.org/aq/. This work has been supported by the Deutsche 
Forschungsgemeinschaft (DFG) through the collaborative research grant SFB 956 
``Conditions and Impact of Star Formation'', project area B3, and through the 
Ger{\"a}tezentrum ``Cologne Center for Terahertz Spectroscopy''. Travel for A.W., D.L., 
and R.V. was partially supported by the French CNRS program ``Physique et Chimie du Milieu 
Interstellaire'' (PCMI). O.H.W. acknowledges support from a Fulbright U.S. Student Research 
Award. R.T.G. is grateful for support from the NASA Astrophysics Theory Program through grant 
NNX11AC38G. Our research benefited from NASA's Astrophysics Data System (ADS).
\end{acknowledgements}


\onecolumn
\begin{appendix}
\label{Appendix}
\section{Complementary figures}

Figures~\ref{f:spec_c3h7cn-n_ve0}--\ref{f:spec_c3h7cn-n-a_v29e1} show the
transitions of \textit{n}-propyl cyanide that are covered by the EMoCA survey 
and contribute significantly to the signal detected toward Sgr~B2(N2).

\begin{figure*}
\centerline{\resizebox{0.82\hsize}{!}{\includegraphics[angle=0]{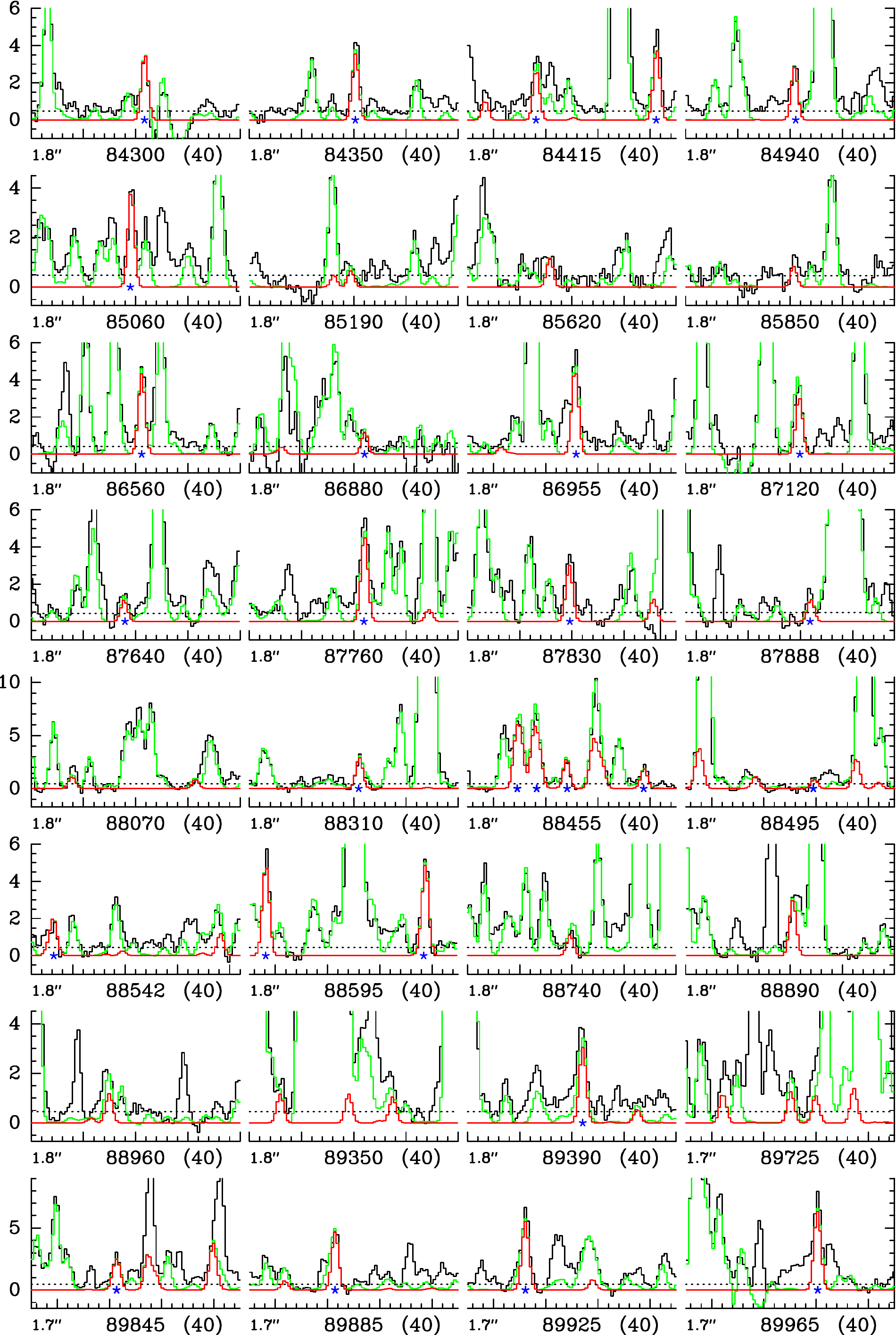}}}
\caption{Transitions of \textit{n-}C$_3$H$_7$CN, $\varv = 0$ covered by our ALMA survey. 
The best-fit LTE synthetic spectrum of \textit{n-}C$_3$H$_7$CN is displayed in red and 
overlaid on the observed spectrum of Sgr~B2(N2) shown in black. The green synthetic 
spectrum contains the contributions of all molecules identified in our survey so far, 
including the species shown in red. 
The central frequency and width are indicated in MHz below each panel. The angular 
resolution (HPBW) is also indicated. The y-axis is labeled in brightness temperature 
units (K). The dotted line indicates the $3\sigma$ noise level. Blue stars indicate 
the lines of \textit{n}-C$_3$H$_7$CN that are counted as detected in Table~\ref{t:ndet}. 
}
\label{f:spec_c3h7cn-n_ve0}
\end{figure*}

\clearpage
\begin{figure*}
\addtocounter{figure}{-1}
\centerline{\resizebox{0.82\hsize}{!}{\includegraphics[angle=0]{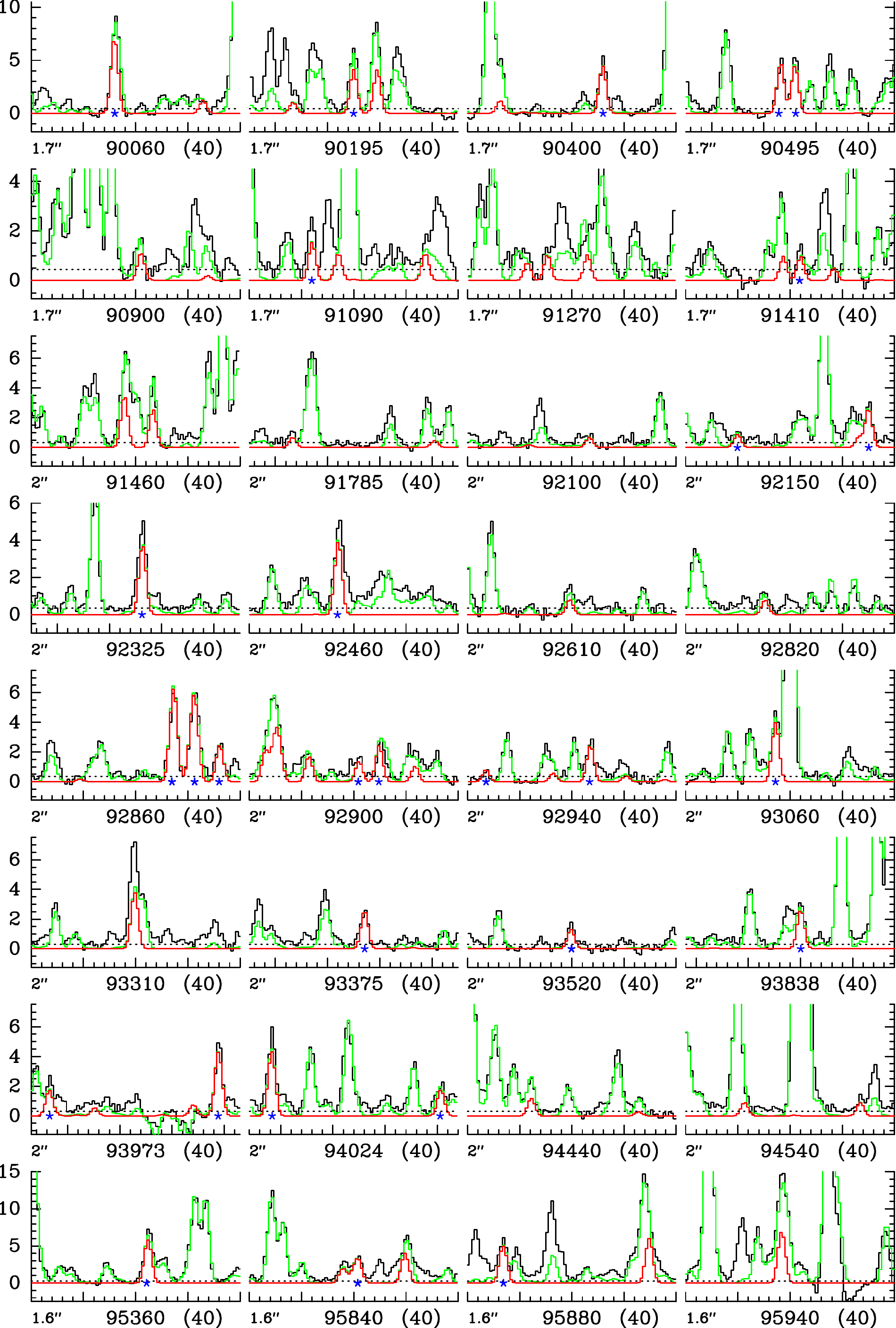}}}
\caption{continued.}
\end{figure*}

\clearpage
\begin{figure*}
\addtocounter{figure}{-1}
\centerline{\resizebox{0.82\hsize}{!}{\includegraphics[angle=0]{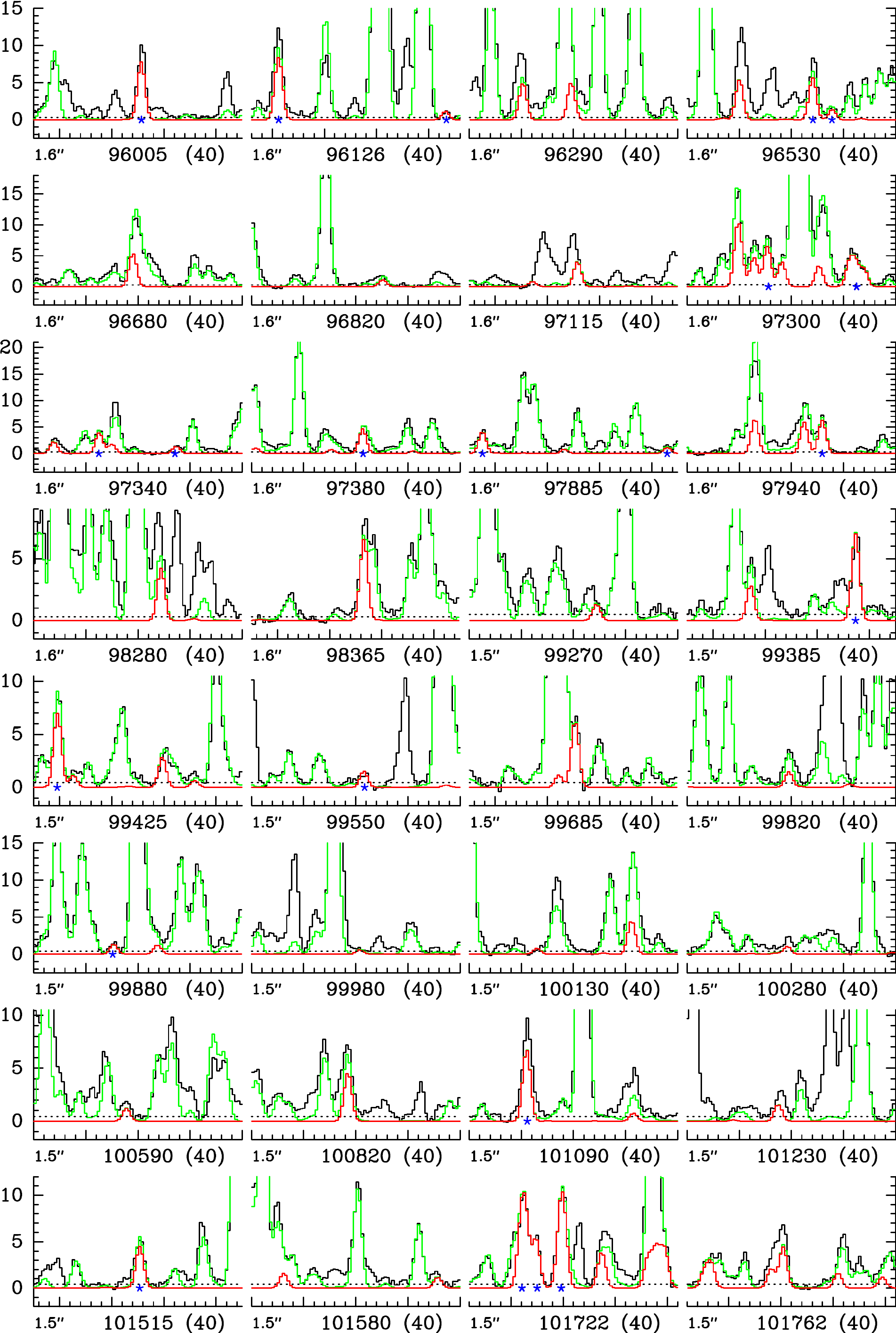}}}
\caption{continued.}
\end{figure*}

\clearpage
\begin{figure*}
\addtocounter{figure}{-1}
\centerline{\resizebox{0.82\hsize}{!}{\includegraphics[angle=0]{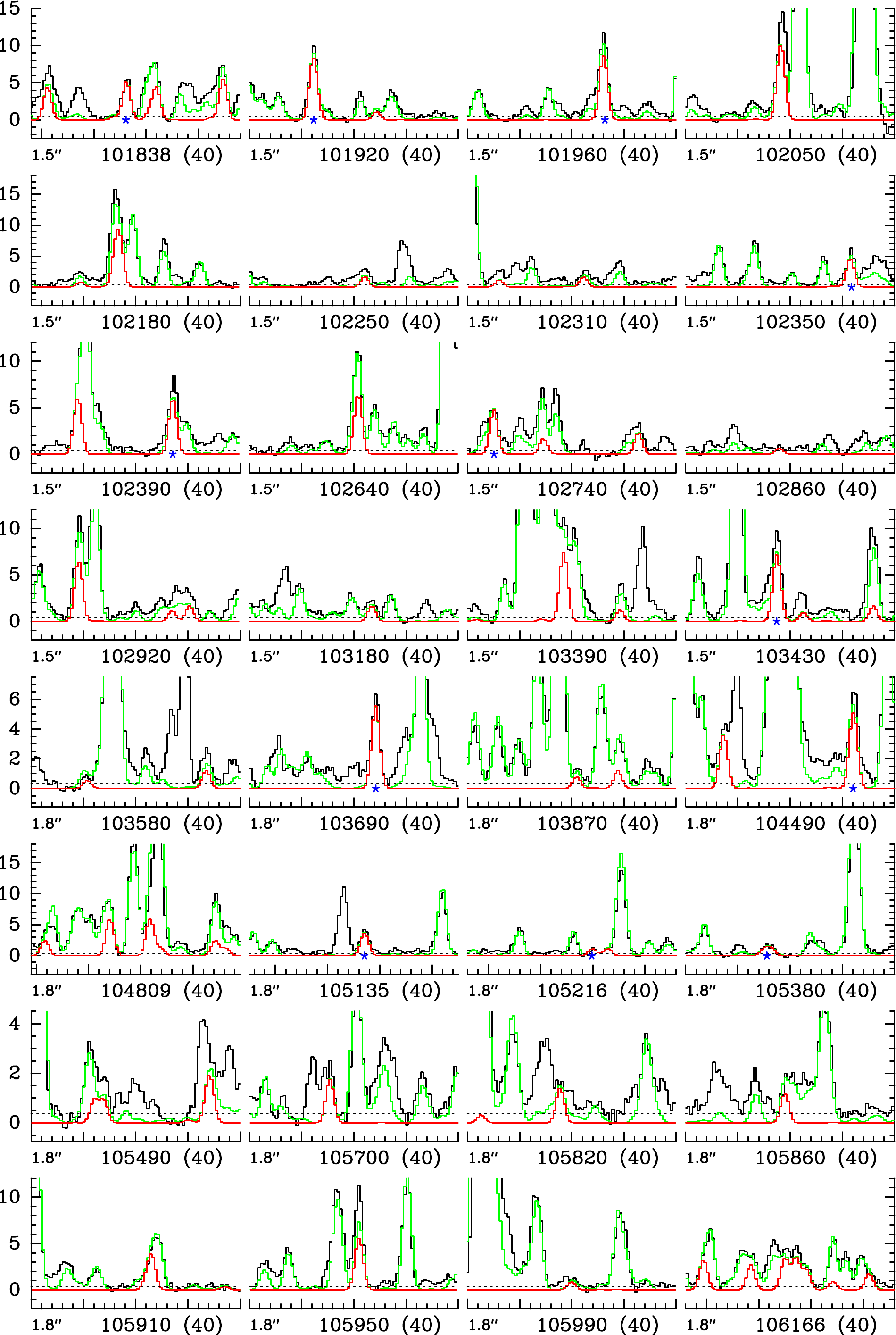}}}
\caption{continued.}
\end{figure*}

\clearpage
\begin{figure*}
\addtocounter{figure}{-1}
\centerline{\resizebox{0.82\hsize}{!}{\includegraphics[angle=0]{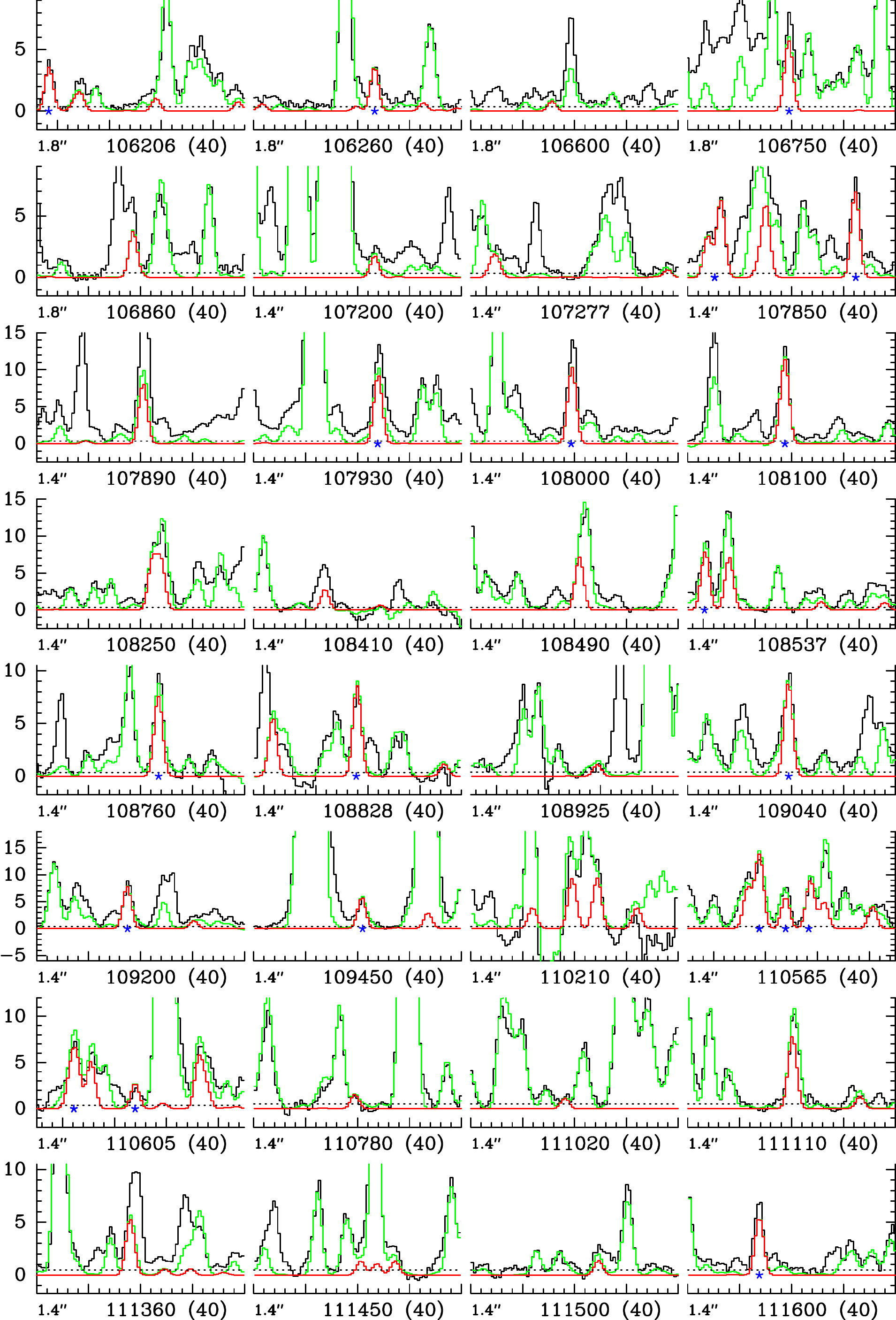}}}
\caption{continued.}
\end{figure*}

\clearpage
\begin{figure*}
\addtocounter{figure}{-1}
\centerline{\resizebox{0.82\hsize}{!}{\includegraphics[angle=0]{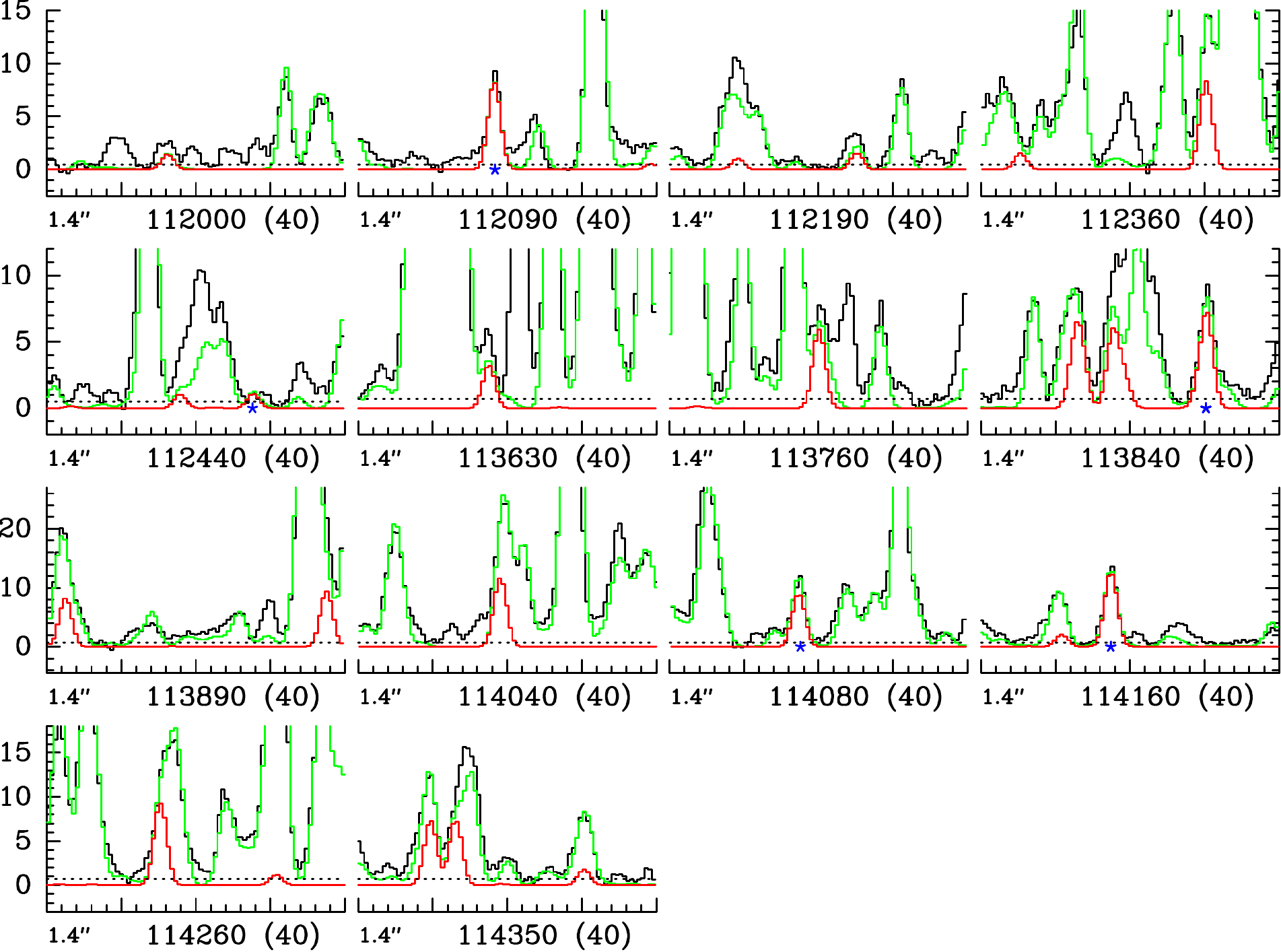}}}
\caption{continued.}
\end{figure*}

\clearpage
\begin{figure*}
\centerline{\resizebox{0.82\hsize}{!}{\includegraphics[angle=0]{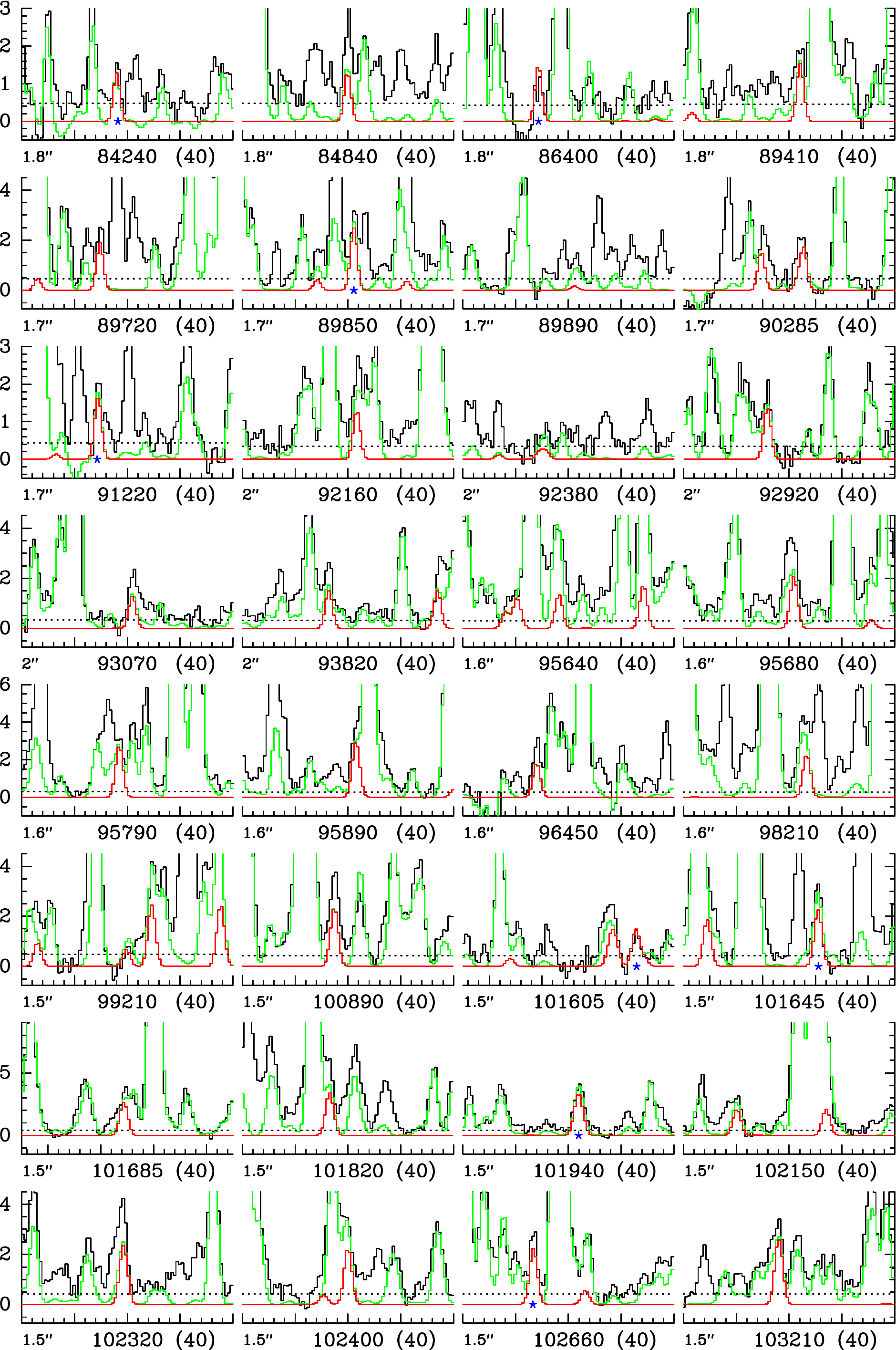}}}
\caption{Same as Fig.~\ref{f:spec_c3h7cn-n_ve0} for \textit{gauche} 
\textit{n-}C$_3$H$_7$CN, $\varv_{30}=1$.
}
\label{f:spec_c3h7cn-n-g_v30e1}
\end{figure*}

\clearpage
\begin{figure*}
\addtocounter{figure}{-1}
\centerline{\resizebox{0.82\hsize}{!}{\includegraphics[angle=0]{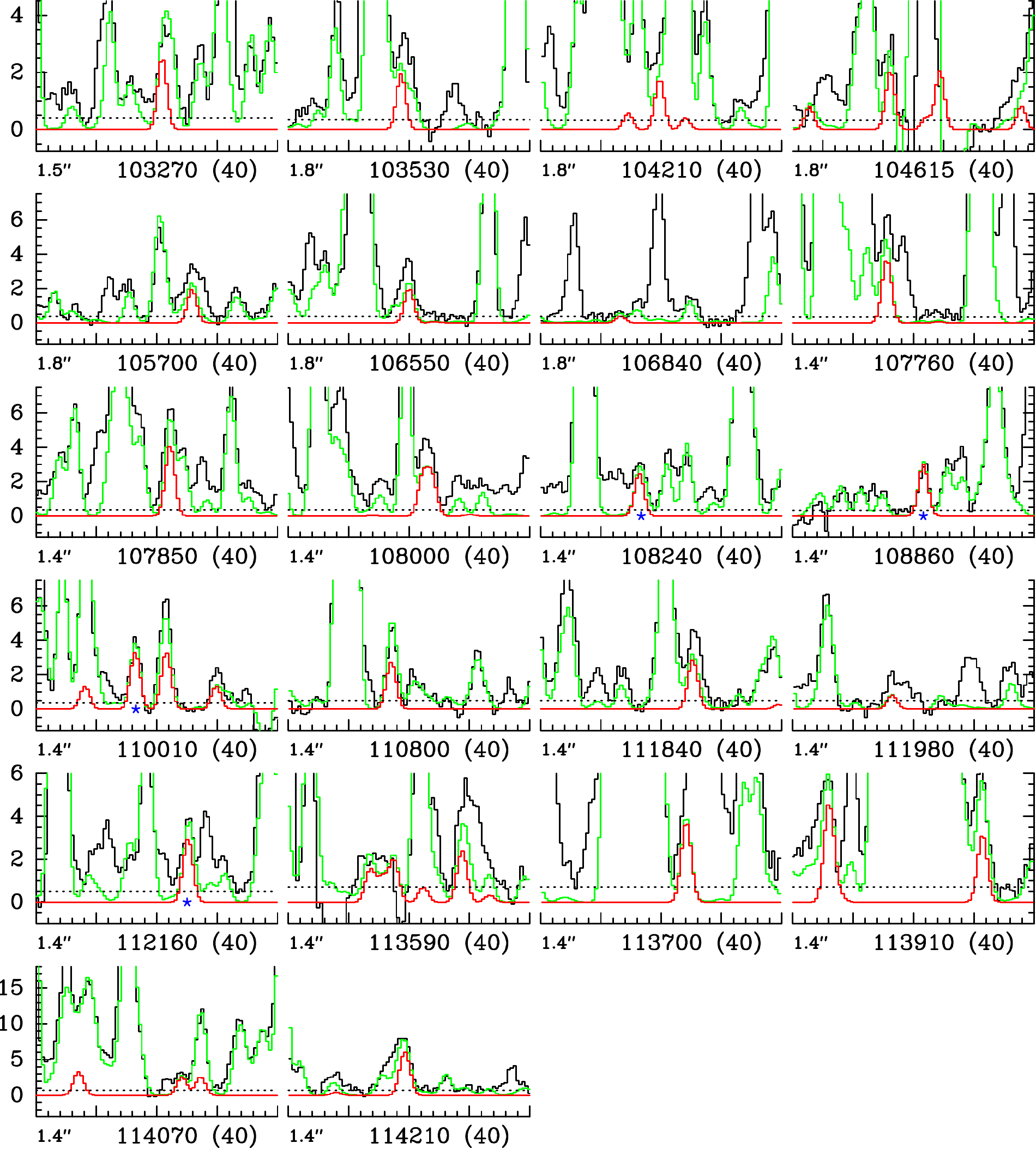}}}
\caption{continued.}
\end{figure*}

\clearpage
\begin{figure*}
\centerline{\resizebox{0.82\hsize}{!}{\includegraphics[angle=0]{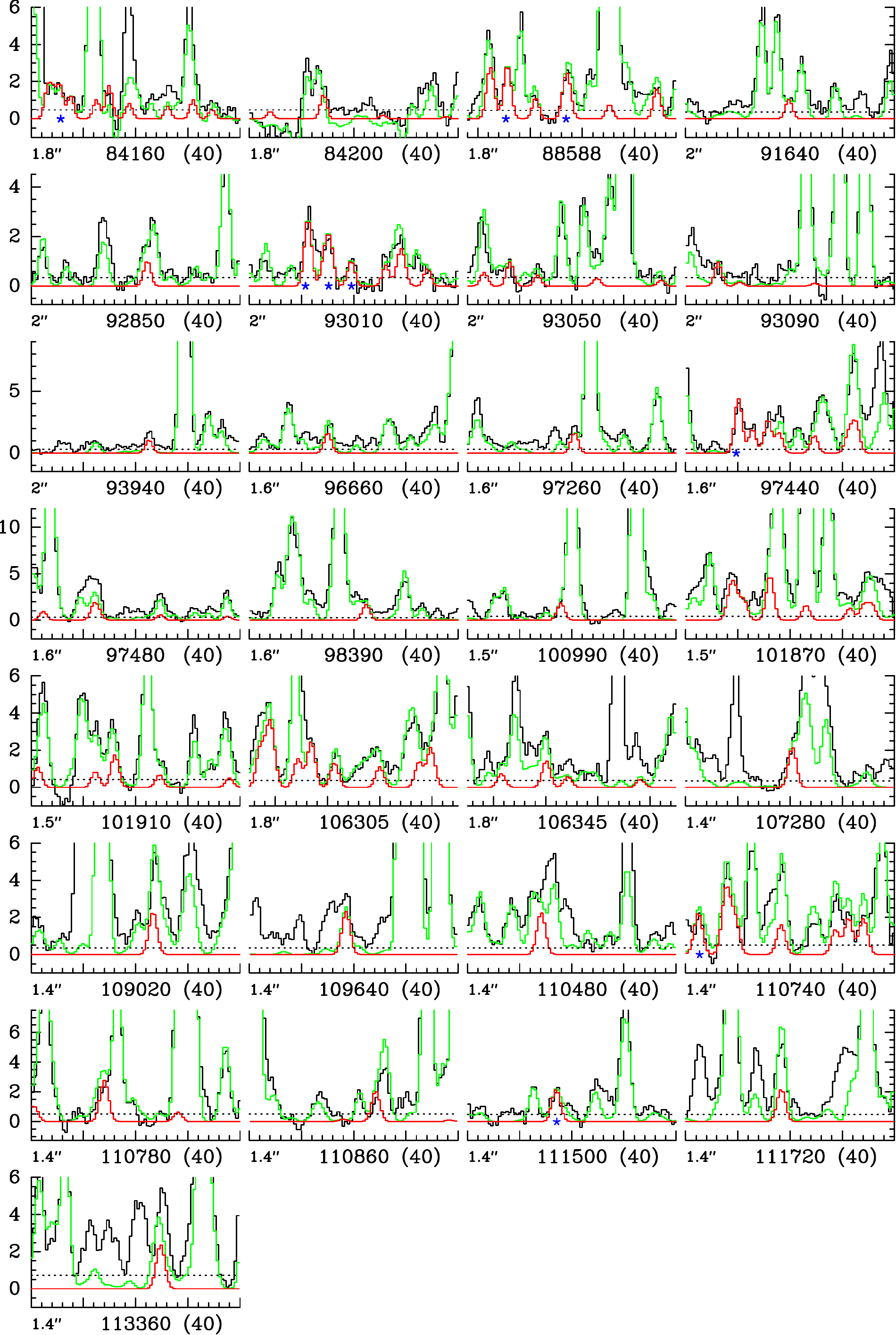}}}
\caption{Same as Fig.~\ref{f:spec_c3h7cn-n_ve0} for \textit{anti} 
\textit{n-}C$_3$H$_7$CN, $\varv_{30}=1$.
}
\label{f:spec_c3h7cn-n-a_v30e1}
\end{figure*}

\clearpage
\begin{figure*}
\centerline{\resizebox{0.82\hsize}{!}{\includegraphics[angle=0]{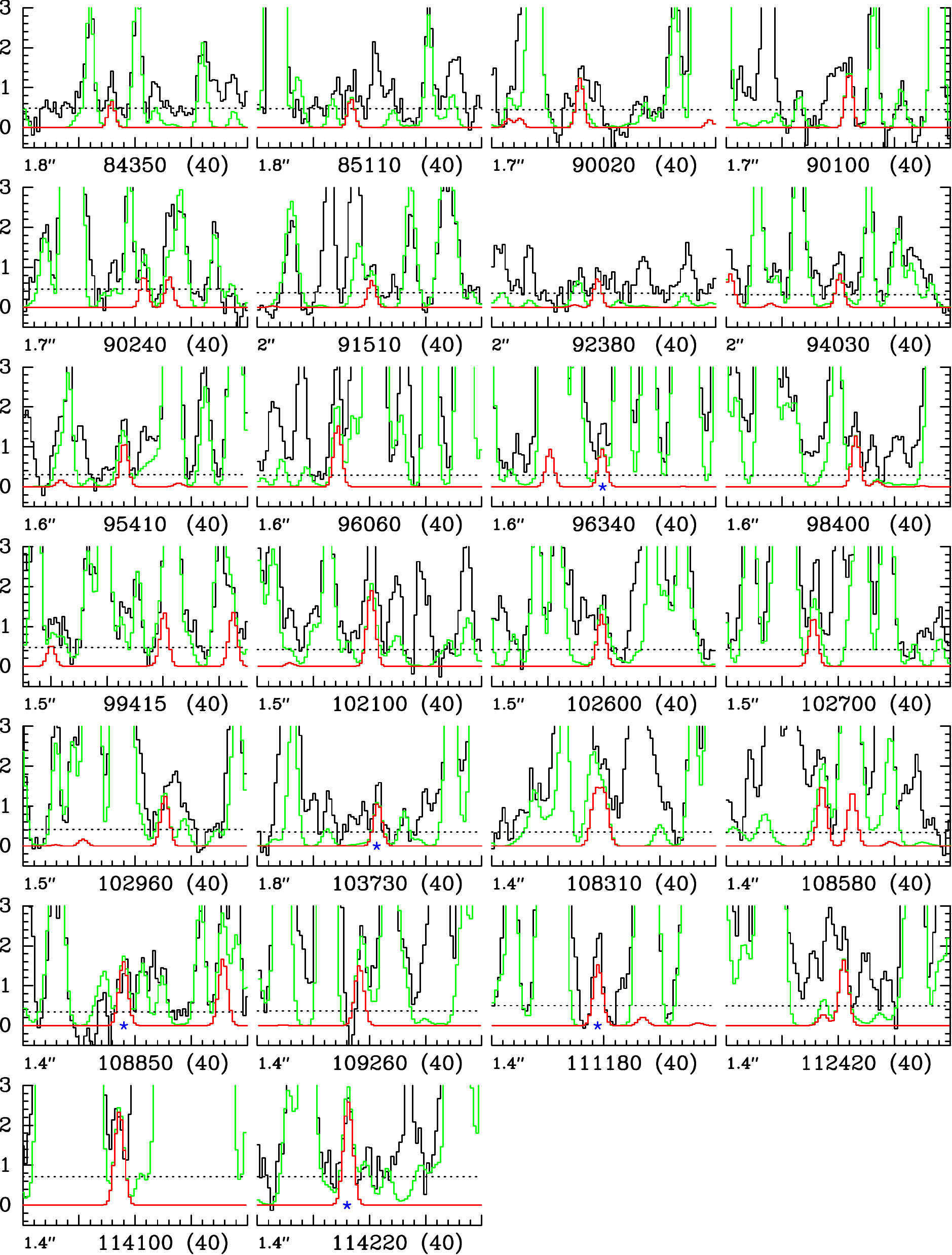}}}
\caption{Same as Fig.~\ref{f:spec_c3h7cn-n_ve0} for \textit{gauche} 
\textit{n-}C$_3$H$_7$CN, $\varv_{29}=1$.
}
\label{f:spec_c3h7cn-n-g_v29e1}
\end{figure*}

\clearpage
\begin{figure*}
\centerline{\resizebox{0.82\hsize}{!}{\includegraphics[angle=0]{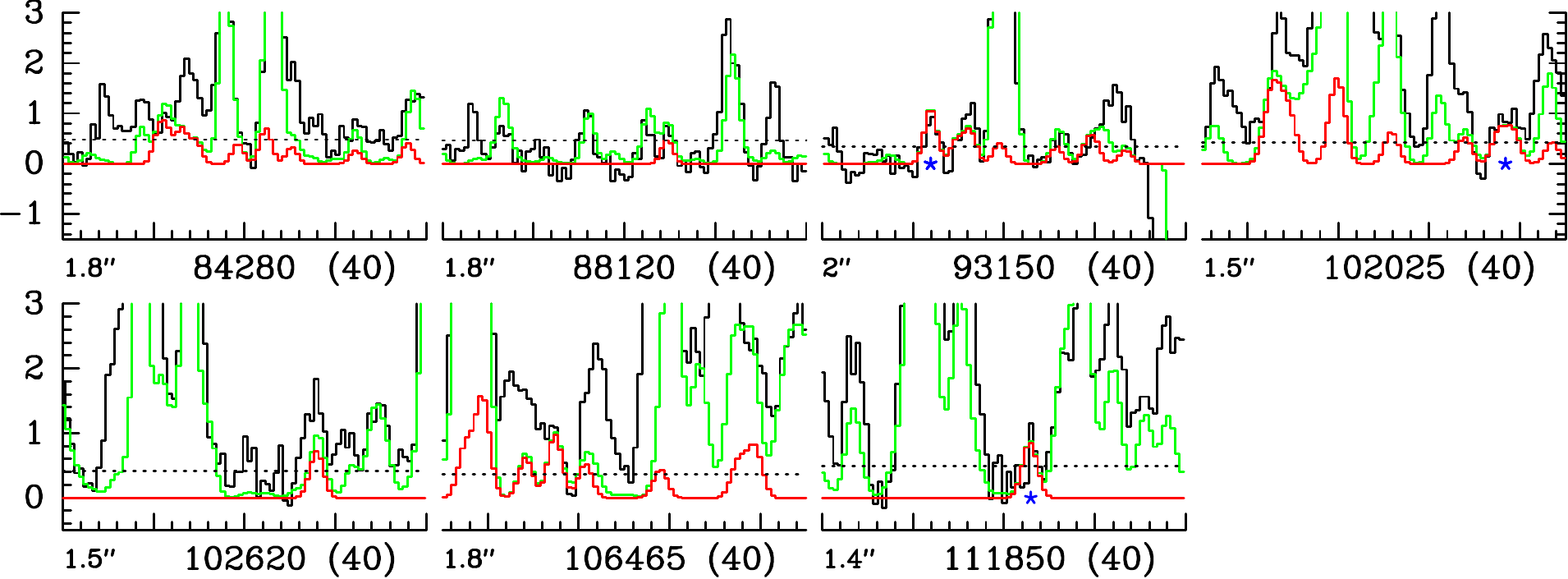}}}
\caption{Same as Fig.~\ref{f:spec_c3h7cn-n_ve0} for \textit{anti} 
\textit{n-}C$_3$H$_7$CN, $\varv_{18}=1$.
}
\label{f:spec_c3h7cn-n-a_v18e1}
\end{figure*}

\clearpage
\begin{figure*}
\centerline{\resizebox{0.82\hsize}{!}{\includegraphics[angle=0]{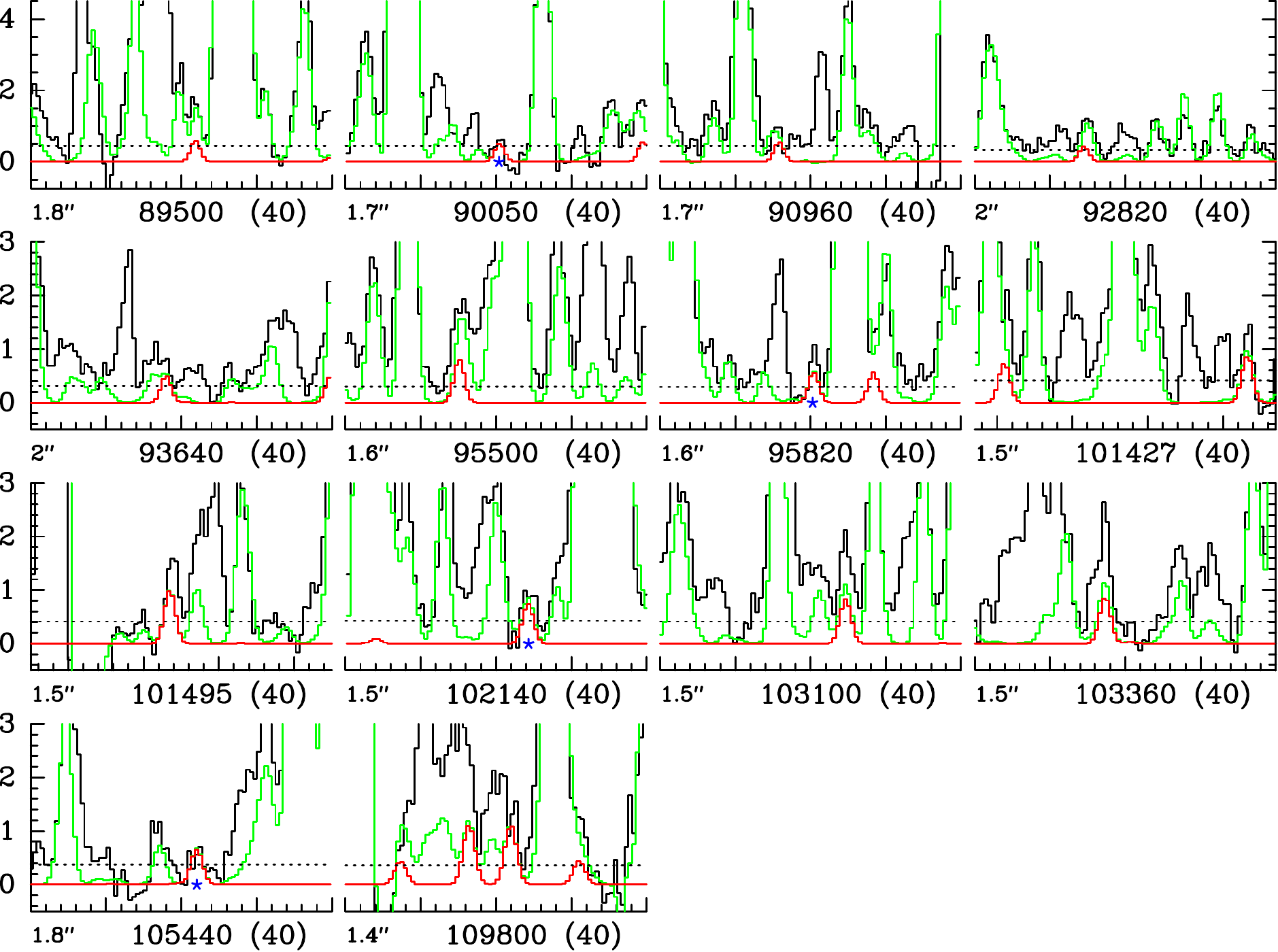}}}
\caption{Same as Fig.~\ref{f:spec_c3h7cn-n_ve0} for \textit{gauche} 
\textit{n-}C$_3$H$_7$CN, $\varv_{30}=2$.
}
\label{f:spec_c3h7cn-n-g_v30e2}
\end{figure*}

\clearpage
\begin{figure*}
\centerline{\resizebox{0.82\hsize}{!}{\includegraphics[angle=0]{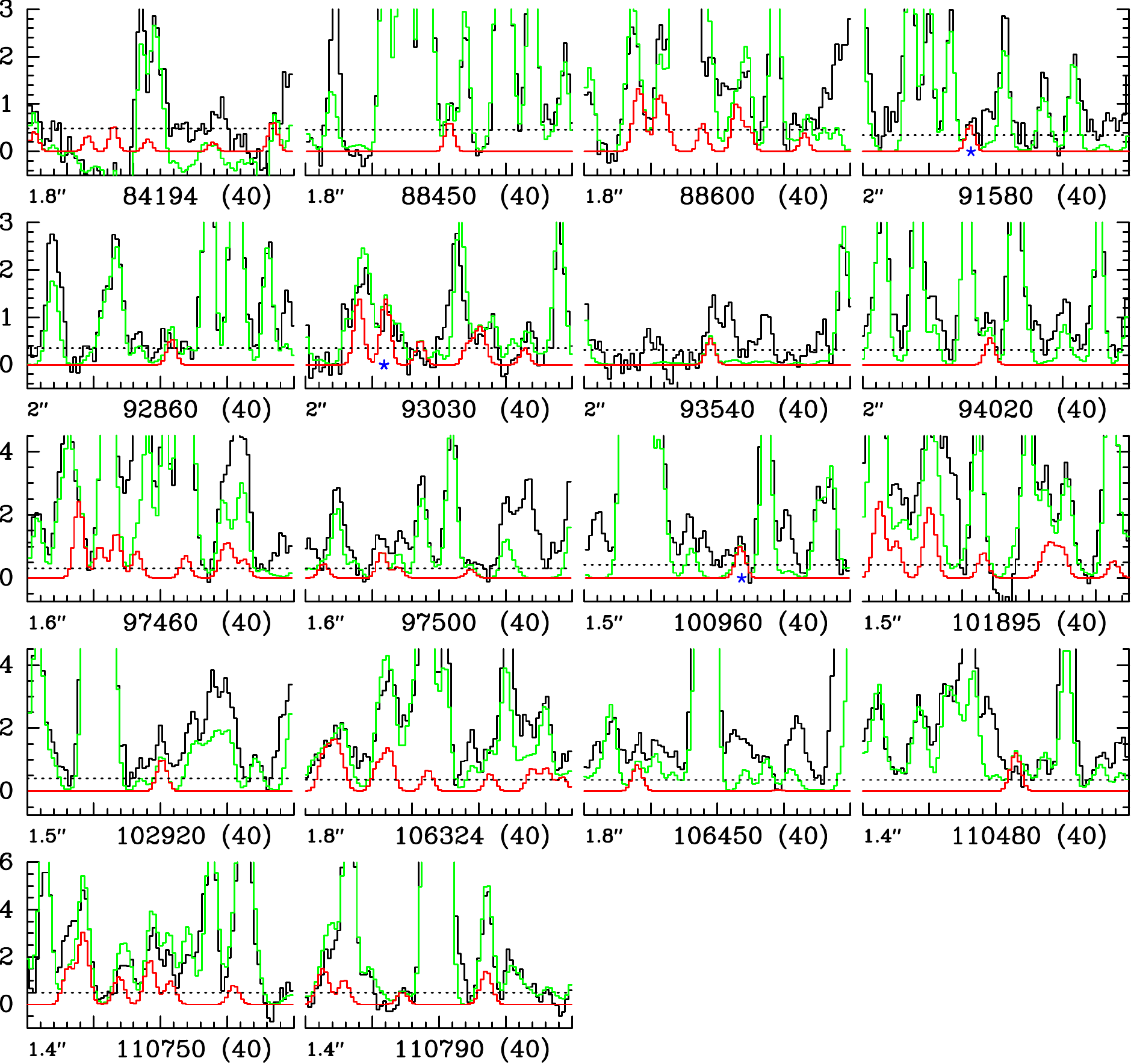}}}
\caption{Same as Fig.~\ref{f:spec_c3h7cn-n_ve0} for \textit{anti} 
\textit{n-}C$_3$H$_7$CN, $\varv_{30}=2$.
}
\label{f:spec_c3h7cn-n-a_v30e2}
\end{figure*}

\clearpage
\begin{figure*}
\centerline{\resizebox{0.82\hsize}{!}{\includegraphics[angle=0]{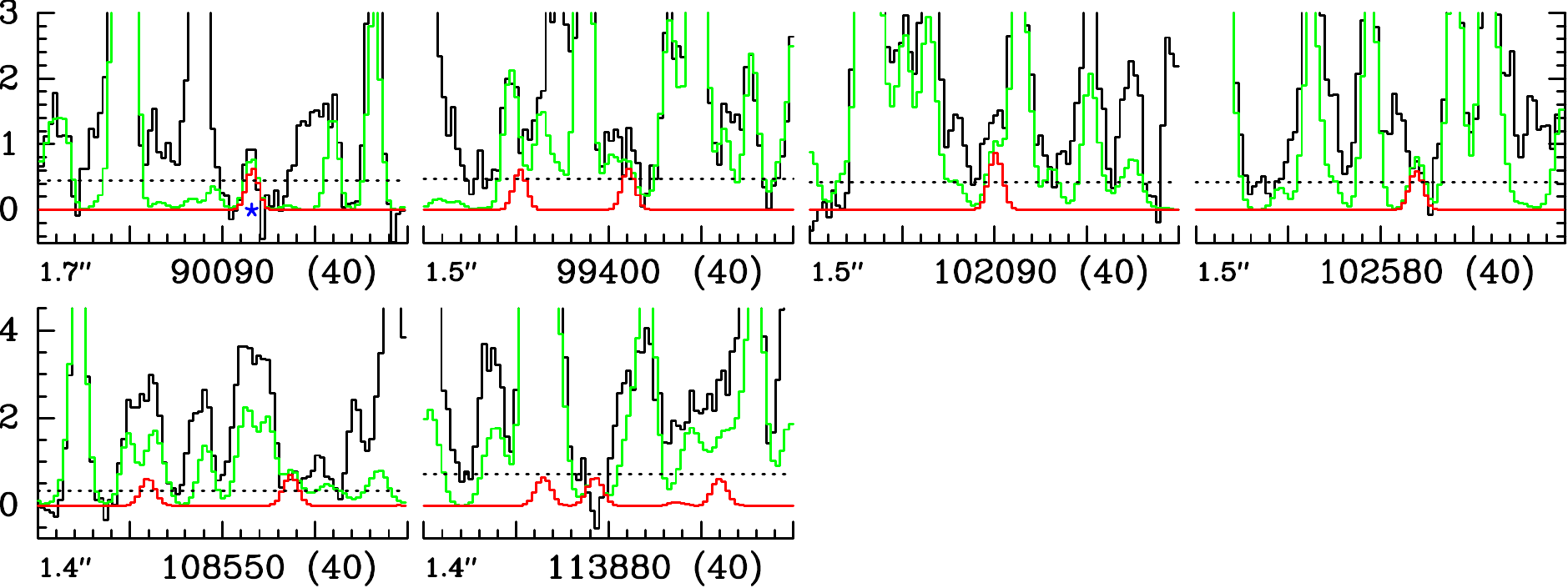}}}
\caption{Same as Fig.~\ref{f:spec_c3h7cn-n_ve0} for \textit{gauche} 
\textit{n-}C$_3$H$_7$CN, $\varv_{28}=1$.
}
\label{f:spec_c3h7cn-n-g_v28e1}
\end{figure*}

\clearpage
\begin{figure*}
\centerline{\resizebox{0.82\hsize}{!}{\includegraphics[angle=0]{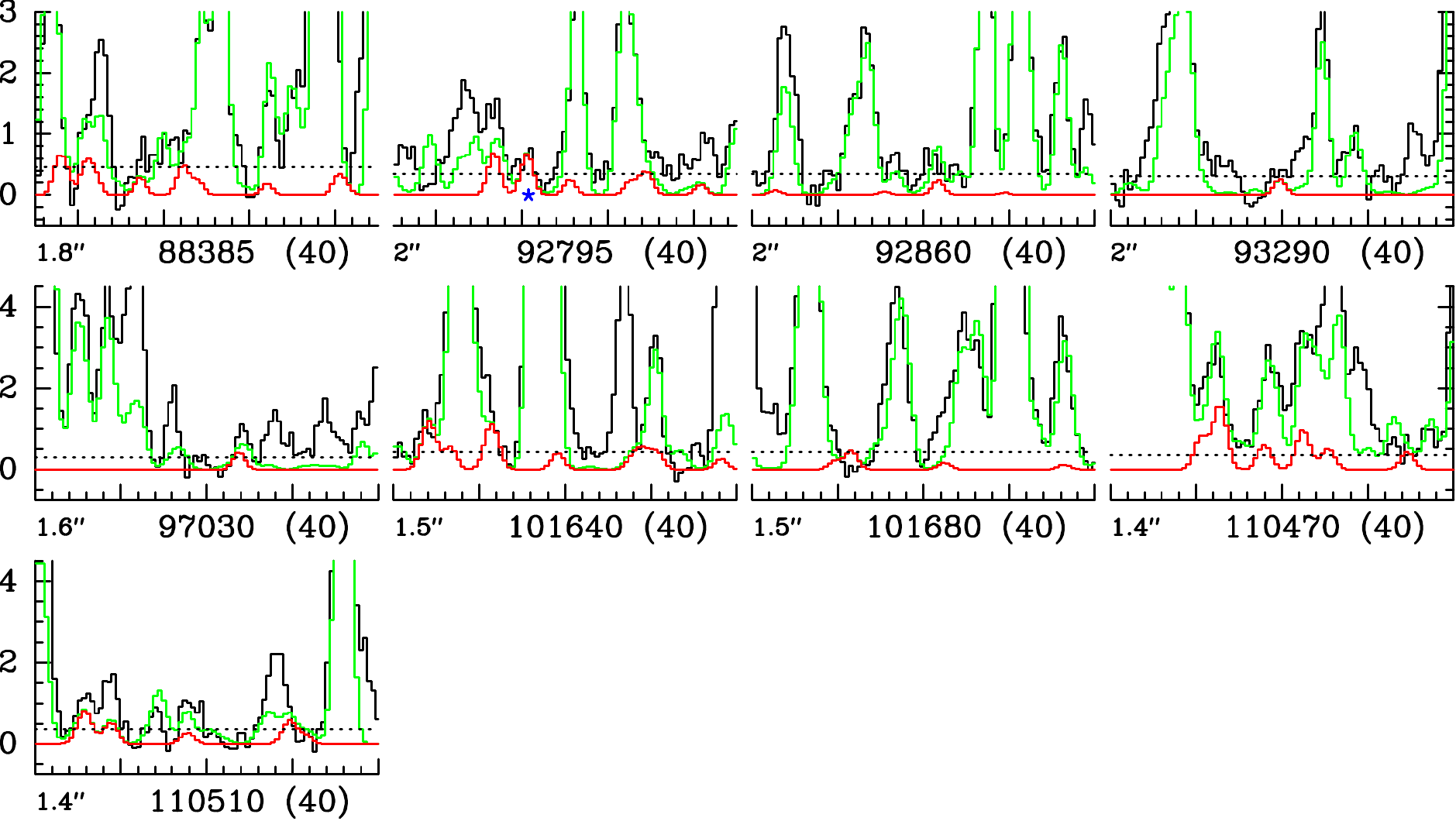}}}
\caption{Same as Fig.~\ref{f:spec_c3h7cn-n_ve0} for \textit{anti} 
\textit{n-}C$_3$H$_7$CN, $\varv_{29}=1$.
}
\label{f:spec_c3h7cn-n-a_v29e1}
\end{figure*}

\section{Complementary tables}

Tables~\ref{t:list_g_v30e1}--\ref{t:list_a_v29e1} list the transitions from 
within vibrationally excited states of \textit{n}-propyl cyanide that are 
detected toward Sgr~B2(N2), meaning that are above the $3\sigma$ level and 
do not suffer much from contamination by other species.

\input{tab_c3h7cn-n-g_v30e1_linelist.tex}
\input{tab_c3h7cn-n-a_v30e1_linelist.tex}
\input{tab_c3h7cn-n-g_v29e1_linelist.tex}
\input{tab_c3h7cn-n-a_v18e1_linelist.tex}
\input{tab_c3h7cn-n-g_v30e2_linelist.tex}
\input{tab_c3h7cn-n-a_v30e2_linelist.tex}
\input{tab_c3h7cn-n-g_v28e1_linelist.tex}
\input{tab_c3h7cn-n-a_v29e1_linelist.tex}

\end{appendix}

\end{document}

%% file: tab_c3h7cn-n_ndet.tex
\begin{table}[!ht]
 {\centering
 \caption{
 Number of lines of \textit{n-}propyl cyanide detected toward Sgr~B2(N2).
}
 \label{t:ndet}
 \vspace*{-1.2ex}
 \begin{tabular}{lrr}
 \hline\hline
 \multicolumn{1}{c}{State} & \multicolumn{1}{c}{$N_{\rm det}$\tablefootmark{a}} & \multicolumn{1}{c}{$N_{\rm trans}$\tablefootmark{b}} \\ 
  & \\ 
 \hline
 \textit{a+g} $\varv=0$ & 116 & 199 \\ 
 \textit{g} $\varv_{30}=1$ & 12 & 16 \\ 
 \textit{a} $\varv_{30}=1$ & 9 & 33 \\ 
 \textit{g} $\varv_{29}=1$ & 6 & 8 \\ 
 \textit{a} $\varv_{18}=1$ & 3 & 7 \\ 
 \textit{g} $\varv_{30}=2$ & 4 & 4 \\ 
 \textit{a} $\varv_{30}=2$ & 3 & 6 \\ 
 \textit{g} $\varv_{28}=1$ & 1 & 2 \\ 
 \textit{a} $\varv_{29}=1$ & 1 & 4 \\ 
\hline 
 \end{tabular}
 }\\[-0.5ex] 
 \tablefoot{
 \tablefoottext{a}{Number of detected lines \citep[conservative estimate, see Sect.~3 of][]{deuterated_SgrB2N2_2016}. One line of a given state may mean a group of transitions of that state that are blended together.}
 \tablefoottext{b}{Number of transitions covered by the detected lines.}
 }
 \end{table}

%% file: tab_c3h7cn-n-g_v30e1_linelist.tex
\begin{table*}
 {\centering
 \caption{
 Selection of lines of \textit{gauche} propyl cyanide $\varv_{30}=1$ covered by the EMoCA survey of Sgr B2(N2).
}
 \label{t:list_g_v30e1}
 \vspace*{0.0ex}
 \begin{tabular}{lrcccrccrrcrr}
 \hline\hline
 \multicolumn{1}{c}{Transition\tablefootmark{a}} & \multicolumn{1}{c}{Frequency} & \multicolumn{1}{c}{Unc.\tablefootmark{b}} & \multicolumn{1}{c}{$E_{\rm up}$\tablefootmark{c}} & \multicolumn{1}{c}{$g_{\rm up}$\tablefootmark{d}} & \multicolumn{1}{c}{$A_{\rm ul}$\tablefootmark{e}} & \multicolumn{1}{c}{$\sigma$\tablefootmark{f}} & \multicolumn{1}{c}{$\tau_{\rm peak}$\tablefootmark{g}} & \multicolumn{2}{c}{Frequency range\tablefootmark{h}} & \multicolumn{1}{c}{$I_{\rm obs}$\tablefootmark{i}} & \multicolumn{1}{c}{$I_{\rm mod}$\tablefootmark{j}} & \multicolumn{1}{c}{$I_{\rm all}$\tablefootmark{k}} \\ 
  & \multicolumn{1}{c}{\scriptsize (MHz)} & \multicolumn{1}{c}{\scriptsize (kHz)} &  \multicolumn{1}{c}{\scriptsize (K)} & & \multicolumn{1}{c}{\scriptsize ($10^{-5}$ s$^{-1}$)} & \multicolumn{1}{c}{\scriptsize (mK)} & & \multicolumn{1}{c}{\scriptsize (MHz)} & \multicolumn{1}{c}{\scriptsize (MHz)} & \multicolumn{1}{c}{\scriptsize (K km s$^{-1}$)} & \multicolumn{2}{c}{\scriptsize (K km s$^{-1}$)} \\ 
 \hline
14$_{4,11}$ -- 13$_{4,10}$ &   84237.836 &   1 &  203 & 58 &  3.3 &  158 &  0.041 &  84236.8 &  84239.7 &   9.1(7)$^\star$ &   6.9 &   6.8 \\ 
14$_{2,12}$ -- 13$_{2,11}$ &   86394.181 &   1 &  199 & 58 &  3.8 &  142 &  0.044 &  86392.9 &  86395.8 &   5.9(6) &   8.0 &   8.1 \\ 
15$_{7,9}$ -- 14$_{7,8}$ &   89850.897 &   1 &  218 & 62 &  3.4 &  149 &  0.073 &  89849.9 &  89852.4 &  19.1(5)$^\star$ &  12.9 &  14.7 \\ 
15$_{7,8}$ -- 14$_{7,7}$ &   89851.023 &   1 &  218 & 62 &  3.4 & -- & -- & -- & -- & -- & -- & -- \\ 
15$_{4,11}$ -- 14$_{4,10}$ &   91214.307 &   1 &  207 & 62 &  4.3 &  149 &  0.047 &  91213.1 &  91215.6 &  10.6(5)$^\star$ &   8.5 &   9.3 \\ 
17$_{13,4}$ -- 16$_{13,3}$ &  101617.584 &   1 &  269 & 70 &  2.6 &  141 &  0.036 & 101616.2 & 101619.6 &   8.2(5)$^\star$ &   9.0 &   9.1 \\ 
17$_{13,5}$ -- 16$_{13,4}$ &  101617.584 &   1 &  269 & 70 &  2.6 & -- & -- & -- & -- & -- & -- & -- \\ 
17$_{11,6}$ -- 16$_{11,5}$ &  101650.360 &   1 &  252 & 70 &  3.7 &  141 &  0.056 & 101648.9 & 101652.3 &  18.5(5)$^\star$ &  12.0 &  17.7 \\ 
17$_{11,7}$ -- 16$_{11,6}$ &  101650.360 &   1 &  252 & 70 &  3.7 & -- & -- & -- & -- & -- & -- & -- \\ 
17$_{7,11}$ -- 16$_{7,10}$ &  101941.402 &   1 &  228 & 70 &  5.3 &  141 &  0.081 & 101939.5 & 101944.4 &  29.6(6)$^\star$ &  20.5 &  26.3 \\ 
17$_{7,10}$ -- 16$_{7,9}$ &  101942.099 &   1 &  228 & 70 &  5.3 & -- & -- & -- & -- & -- & -- & -- \\ 
17$_{5,12}$ -- 16$_{5,11}$ &  102653.203 &   1 &  220 & 70 &  6.0 &  141 &  0.055 & 102651.6 & 102655.0 &  18.5(5)$^\star$ &  12.0 &  12.7 \\ 
18$_{6,13}$ -- 17$_{6,12}$ &  108236.201 &   1 &  228 & 74 &  6.8 &  115 &  0.056 & 108235.3 & 108238.2 &  20.7(4)$^\star$ &  12.4 &  16.9 \\ 
19$_{1,18}$ -- 18$_{1,17}$ &  108861.471 &   1 &  221 & 78 &  7.7 &  115 &  0.070 & 108859.5 & 108863.9 &  18.3(5)$^\star$ &  16.6 &  17.6 \\ 
20$_{1,20}$ -- 19$_{1,19}$ &  110006.258 &   1 &  223 & 82 &  8.0 &  123 &  0.074 & 110004.6 & 110008.5 &  22.2(5)$^\star$ &  17.9 &  20.7 \\ 
19$_{3,17}$ -- 18$_{3,16}$ &  112165.026 &   1 &  224 & 78 &  8.3 &  166 &  0.068 & 112163.4 & 112166.8 &  25.3(6)$^\star$ &  15.6 &  21.4 \\ 
 \hline
 \end{tabular}
 }\\[1ex] 
 \tablefoot{
 \tablefoottext{a}{Quantum numbers of the upper and lower levels.}
 \tablefoottext{b}{Frequency uncertainty.}
 \tablefoottext{c}{Upper level energy.}
 \tablefoottext{d}{Upper level degeneracy.}
 \tablefoottext{e}{Einstein coefficient for spontaneous emission.}
 \tablefoottext{f}{Measured rms noise level.}
 \tablefoottext{g}{Peak opacity of the synthetic line.}
 \tablefoottext{h}{Frequency range over which the emission was integrated.}
 \tablefoottext{i}{Integrated intensity of the observed spectrum in brightness temperature scale. The statistical standard deviation is given in parentheses in unit of the last digit. Values marked with a star are used in the population diagram shown in Fig.~\ref{f:popdiag}.}
 \tablefoottext{j}{Integrated intensity of the synthetic spectrum of the selected state.}
 \tablefoottext{k}{Integrated intensity of the model that contains the contribution of all identified molecules, including \textit{n}-propyl cyanide and its vibrationally excited states.}
 }
 \end{table*}

%% file: tab_c3h7cn-n-a_v30e1_linelist.tex
\begin{table*}
 {\centering
 \caption{
 Selection of lines of \textit{anti} propyl cyanide $\varv_{30}=1$ covered by the EMoCA survey of Sgr B2(N2).
}
 \label{t:list_a_v30e1}
 \vspace*{0.0ex}
 \begin{tabular}{lrcccrccrrcrr}
 \hline\hline
 \multicolumn{1}{c}{Transition\tablefootmark{a}} & \multicolumn{1}{c}{Frequency} & \multicolumn{1}{c}{Unc.\tablefootmark{b}} & \multicolumn{1}{c}{$E_{\rm up}$\tablefootmark{c}} & \multicolumn{1}{c}{$g_{\rm up}$\tablefootmark{d}} & \multicolumn{1}{c}{$A_{\rm ul}$\tablefootmark{e}} & \multicolumn{1}{c}{$\sigma$\tablefootmark{f}} & \multicolumn{1}{c}{$\tau_{\rm peak}$\tablefootmark{g}} & \multicolumn{2}{c}{Frequency range\tablefootmark{h}} & \multicolumn{1}{c}{$I_{\rm obs}$\tablefootmark{i}} & \multicolumn{1}{c}{$I_{\rm mod}$\tablefootmark{j}} & \multicolumn{1}{c}{$I_{\rm all}$\tablefootmark{k}} \\ 
  & \multicolumn{1}{c}{\scriptsize (MHz)} & \multicolumn{1}{c}{\scriptsize (kHz)} &  \multicolumn{1}{c}{\scriptsize (K)} & & \multicolumn{1}{c}{\scriptsize ($10^{-5}$ s$^{-1}$)} & \multicolumn{1}{c}{\scriptsize (mK)} & & \multicolumn{1}{c}{\scriptsize (MHz)} & \multicolumn{1}{c}{\scriptsize (MHz)} & \multicolumn{1}{c}{\scriptsize (K km s$^{-1}$)} & \multicolumn{2}{c}{\scriptsize (K km s$^{-1}$)} \\ 
 \hline
19$_{6,14}$ -- 18$_{6,13}$ &   84142.973 &   1 &  276 & 39 &  4.9 &  158 &  0.063 &  84142.5 &  84148.8 &  30.7(10) &  30.5 &  29.0 \\ 
19$_{6,13}$ -- 18$_{6,12}$ &   84142.973 &   1 &  276 & 39 &  4.9 & -- & -- & -- & -- & -- & -- & -- \\ 
19$_{7,12}$ -- 18$_{7,11}$ &   84144.088 &   1 &  289 & 39 &  4.7 & -- & -- & -- & -- & -- & -- & -- \\ 
19$_{7,13}$ -- 18$_{7,12}$ &   84144.088 &   1 &  289 & 39 &  4.7 & -- & -- & -- & -- & -- & -- & -- \\ 
19$_{5,15}$ -- 18$_{5,14}$ &   84145.416 &   1 &  265 & 39 &  5.0 & -- & -- & -- & -- & -- & -- & -- \\ 
19$_{5,14}$ -- 18$_{5,13}$ &   84145.420 &   1 &  265 & 39 &  5.0 & -- & -- & -- & -- & -- & -- & -- \\ 
19$_{8,11}$ -- 18$_{8,10}$ &   84147.434 &   1 &  304 & 39 &  4.5 & -- & -- & -- & -- & -- & -- & -- \\ 
19$_{8,12}$ -- 18$_{8,11}$ &   84147.434 &   1 &  304 & 39 &  4.5 & -- & -- & -- & -- & -- & -- & -- \\ 
20$_{5,16}$ -- 19$_{5,15}$ &   88575.296 &   1 &  269 & 41 &  5.9 &  153 &  0.088 &  88574.2 &  88576.7 &  16.9(6) &  16.0 &  16.6 \\ 
20$_{5,15}$ -- 19$_{5,14}$ &   88575.302 &   1 &  269 & 41 &  5.9 & -- & -- & -- & -- & -- & -- & -- \\ 
20$_{8,12}$ -- 19$_{8,11}$ &   88575.904 &   1 &  308 & 41 &  5.3 & -- & -- & -- & -- & -- & -- & -- \\ 
20$_{8,13}$ -- 19$_{8,12}$ &   88575.904 &   1 &  308 & 41 &  5.3 & -- & -- & -- & -- & -- & -- & -- \\ 
20$_{4,17}$ -- 19$_{4,16}$ &   88586.378 &   1 &  261 & 41 &  6.1 &  153 &  0.079 &  88585.0 &  88588.9 &  21.6(7) &  16.4 &  23.7 \\ 
20$_{4,16}$ -- 19$_{4,15}$ &   88586.967 &   1 &  261 & 41 &  6.1 & -- & -- & -- & -- & -- & -- & -- \\ 
20$_{10,10}$ -- 19$_{10,9}$ &   88587.381 &   1 &  344 & 41 &  4.7 & -- & -- & -- & -- & -- & -- & -- \\ 
20$_{10,11}$ -- 19$_{10,10}$ &   88587.381 &   1 &  344 & 41 &  4.7 & -- & -- & -- & -- & -- & -- & -- \\ 
21$_{6,16}$ -- 20$_{6,15}$ &   93001.000 &   1 &  285 & 43 &  6.7 &  117 &  0.104 &  92999.3 &  93002.2 &  17.6(5)$^\star$ &  14.4 &  14.6 \\ 
21$_{6,15}$ -- 20$_{6,14}$ &   93001.000 &   1 &  285 & 43 &  6.7 & -- & -- & -- & -- & -- & -- & -- \\ 
21$_{7,14}$ -- 20$_{7,13}$ &   93001.254 &   1 &  298 & 43 &  6.5 & -- & -- & -- & -- & -- & -- & -- \\ 
21$_{7,15}$ -- 20$_{7,14}$ &   93001.254 &   1 &  298 & 43 &  6.5 & -- & -- & -- & -- & -- & -- & -- \\ 
21$_{8,13}$ -- 20$_{8,12}$ &   93004.322 &   1 &  313 & 43 &  6.3 &  117 &  0.083 &  93003.2 &  93007.1 &  15.4(5) &  15.4 &  16.3 \\ 
21$_{8,14}$ -- 20$_{8,13}$ &   93004.322 &   1 &  313 & 43 &  6.3 & -- & -- & -- & -- & -- & -- & -- \\ 
21$_{5,17}$ -- 20$_{5,16}$ &   93005.356 &   1 &  274 & 43 &  6.9 & -- & -- & -- & -- & -- & -- & -- \\ 
21$_{5,16}$ -- 20$_{5,15}$ &   93005.365 &   1 &  274 & 43 &  6.9 & -- & -- & -- & -- & -- & -- & -- \\ 
21$_{9,12}$ -- 20$_{9,11}$ &   93009.378 &   1 &  329 & 43 &  6.0 &  117 &  0.038 &  93008.1 &  93011.0 &   6.0(5)$^\star$ &   5.4 &   6.0 \\ 
21$_{9,13}$ -- 20$_{9,12}$ &   93009.378 &   1 &  329 & 43 &  6.0 & -- & -- & -- & -- & -- & -- & -- \\ 
22$_{7,15}$ -- 21$_{7,14}$ &   97429.823 &   1 &  302 & 45 &  7.6 &  100 &  0.119 &  97428.0 &  97431.4 &  23.4(4)$^\star$ &  24.0 &  24.1 \\ 
22$_{7,16}$ -- 21$_{7,15}$ &   97429.823 &   1 &  302 & 45 &  7.6 & -- & -- & -- & -- & -- & -- & -- \\ 
22$_{6,17}$ -- 21$_{6,16}$ &   97430.109 &   1 &  290 & 45 &  7.8 & -- & -- & -- & -- & -- & -- & -- \\ 
22$_{6,16}$ -- 21$_{6,15}$ &   97430.109 &   1 &  290 & 45 &  7.8 & -- & -- & -- & -- & -- & -- & -- \\ 
25$_{9,16}$ -- 24$_{9,15}$ &  110722.260 &   1 &  349 & 51 & 10.8 &  166 &  0.051 & 110721.0 & 110724.5 &  11.8(6)$^\star$ &  11.6 &  16.2 \\ 
25$_{9,17}$ -- 24$_{9,16}$ &  110722.260 &   1 &  349 & 51 & 10.8 & -- & -- & -- & -- & -- & -- & -- \\ 
25$_{2,23}$ -- 24$_{2,22}$ &  111496.841 &   2 &  273 & 51 & 12.6 &  166 &  0.050 & 111495.2 & 111499.1 &  14.5(6)$^\star$ &  11.5 &  14.9 \\ 
 \hline
 \end{tabular}
 }\\[1ex] 
 \tablefoot{
 See Table~\ref{t:list_g_v30e1}.
 }
 \end{table*}

%% file: tab_c3h7cn-n-g_v29e1_linelist.tex
\begin{table*}
 {\centering
 \caption{
 Selection of lines of \textit{gauche} propyl cyanide $\varv_{29}=1$ covered by the EMoCA survey of Sgr B2(N2).
}
 \label{t:list_g_v29e1}
 \vspace*{0.0ex}
 \begin{tabular}{lrcccrccrrcrr}
 \hline\hline
 \multicolumn{1}{c}{Transition\tablefootmark{a}} & \multicolumn{1}{c}{Frequency} & \multicolumn{1}{c}{Unc.\tablefootmark{b}} & \multicolumn{1}{c}{$E_{\rm up}$\tablefootmark{c}} & \multicolumn{1}{c}{$g_{\rm up}$\tablefootmark{d}} & \multicolumn{1}{c}{$A_{\rm ul}$\tablefootmark{e}} & \multicolumn{1}{c}{$\sigma$\tablefootmark{f}} & \multicolumn{1}{c}{$\tau_{\rm peak}$\tablefootmark{g}} & \multicolumn{2}{c}{Frequency range\tablefootmark{h}} & \multicolumn{1}{c}{$I_{\rm obs}$\tablefootmark{i}} & \multicolumn{1}{c}{$I_{\rm mod}$\tablefootmark{j}} & \multicolumn{1}{c}{$I_{\rm all}$\tablefootmark{k}} \\ 
  & \multicolumn{1}{c}{\scriptsize (MHz)} & \multicolumn{1}{c}{\scriptsize (kHz)} &  \multicolumn{1}{c}{\scriptsize (K)} & & \multicolumn{1}{c}{\scriptsize ($10^{-5}$ s$^{-1}$)} & \multicolumn{1}{c}{\scriptsize (mK)} & & \multicolumn{1}{c}{\scriptsize (MHz)} & \multicolumn{1}{c}{\scriptsize (MHz)} & \multicolumn{1}{c}{\scriptsize (K km s$^{-1}$)} & \multicolumn{2}{c}{\scriptsize (K km s$^{-1}$)} \\ 
 \hline
16$_{6,10}$ -- 15$_{6,9}$ &   96339.593 &   1 &  309 & 66 &  4.6 &  155 &  0.025 &  96338.6 &  96341.0 &   5.7(5)$^\star$ &   4.8 &   5.2 \\ 
18$_{1,17}$ -- 17$_{1,16}$ &  103731.332 &   2 &  307 & 74 &  6.6 &  114 &  0.035 & 103729.6 & 103733.1 &   8.6(4)$^\star$ &   5.7 &   6.2 \\ 
15$_{8,7}$ -- 15$_{7,8}$ &  105942.286 &   3 &  314 & 62 &  1.3 &  123 &  0.011 & 105941.0 & 105943.9 &   2.2(4)$^\star$ &   1.7 &   2.5 \\ 
15$_{8,8}$ -- 15$_{7,9}$ &  105942.502 &   3 &  314 & 62 &  1.3 & -- & -- & -- & -- & -- & -- & -- \\ 
18$_{2,16}$ -- 17$_{2,15}$ &  108847.764 &   2 &  309 & 74 &  7.6 &  115 &  0.036 & 108846.3 & 108849.7 &  10.4(4)$^\star$ &   8.6 &   9.9 \\ 
18$_{4,14}$ -- 17$_{4,13}$ &  111178.807 &   2 &  313 & 74 &  7.9 &  166 &  0.036 & 111177.2 & 111180.6 &  11.9(6)$^\star$ &   8.0 &   8.1 \\ 
19$_{8,12}$ -- 18$_{8,11}$ &  114215.993 &   2 &  334 & 78 &  7.4 &  242 &  0.058 & 114214.1 & 114218.0 &  16.7(9)$^\star$ &  14.0 &  17.4 \\ 
19$_{8,11}$ -- 18$_{8,10}$ &  114216.118 &   2 &  334 & 78 &  7.4 & -- & -- & -- & -- & -- & -- & -- \\ 
 \hline
 \end{tabular}
 }\\[1ex] 
 \tablefoot{
 See Table~\ref{t:list_g_v30e1}.
 }
 \end{table*}

%% file: tab_c3h7cn-n-a_v18e1_linelist.tex
\begin{table*}
 {\centering
 \caption{
 Selection of lines of \textit{anti} propyl cyanide $\varv_{18}=1$ covered by the EMoCA survey of Sgr B2(N2).
}
 \label{t:list_a_v18e1}
 \vspace*{0.0ex}
 \begin{tabular}{lrcccrccrrcrr}
 \hline\hline
 \multicolumn{1}{c}{Transition\tablefootmark{a}} & \multicolumn{1}{c}{Frequency} & \multicolumn{1}{c}{Unc.\tablefootmark{b}} & \multicolumn{1}{c}{$E_{\rm up}$\tablefootmark{c}} & \multicolumn{1}{c}{$g_{\rm up}$\tablefootmark{d}} & \multicolumn{1}{c}{$A_{\rm ul}$\tablefootmark{e}} & \multicolumn{1}{c}{$\sigma$\tablefootmark{f}} & \multicolumn{1}{c}{$\tau_{\rm peak}$\tablefootmark{g}} & \multicolumn{2}{c}{Frequency range\tablefootmark{h}} & \multicolumn{1}{c}{$I_{\rm obs}$\tablefootmark{i}} & \multicolumn{1}{c}{$I_{\rm mod}$\tablefootmark{j}} & \multicolumn{1}{c}{$I_{\rm all}$\tablefootmark{k}} \\ 
  & \multicolumn{1}{c}{\scriptsize (MHz)} & \multicolumn{1}{c}{\scriptsize (kHz)} &  \multicolumn{1}{c}{\scriptsize (K)} & & \multicolumn{1}{c}{\scriptsize ($10^{-5}$ s$^{-1}$)} & \multicolumn{1}{c}{\scriptsize (mK)} & & \multicolumn{1}{c}{\scriptsize (MHz)} & \multicolumn{1}{c}{\scriptsize (MHz)} & \multicolumn{1}{c}{\scriptsize (K km s$^{-1}$)} & \multicolumn{2}{c}{\scriptsize (K km s$^{-1}$)} \\ 
 \hline
21$_{6,16}$ -- 20$_{6,15}$ &   93141.752 &   1 &  426 & 43 &  6.7 &  117 &  0.041 &  93140.5 &  93143.4 &   5.4(5)$^\star$ &   6.1 &   6.1 \\ 
21$_{6,15}$ -- 20$_{6,14}$ &   93141.752 &   1 &  426 & 43 &  6.7 & -- & -- & -- & -- & -- & -- & -- \\ 
21$_{7,14}$ -- 20$_{7,13}$ &   93141.838 &   1 &  439 & 43 &  6.5 & -- & -- & -- & -- & -- & -- & -- \\ 
21$_{7,15}$ -- 20$_{7,14}$ &   93141.838 &   1 &  439 & 43 &  6.5 & -- & -- & -- & -- & -- & -- & -- \\ 
23$_{4,20}$ -- 22$_{4,19}$ &  102037.514 &   1 &  417 & 47 &  9.4 &  141 &  0.018 & 102036.2 & 102040.6 &   9.5(6)$^\star$ &   6.9 &   7.2 \\ 
23$_{4,19}$ -- 22$_{4,18}$ &  102039.055 &   1 &  417 & 47 &  9.4 & -- & -- & -- & -- & -- & -- & -- \\ 
25$_{1,24}$ -- 24$_{1,23}$ &  111852.657 &   2 &  413 & 51 & 12.7 &  166 &  0.020 & 111851.3 & 111854.7 &   4.6(6)$^\star$ &   4.5 &   4.8 \\ 
 \hline
 \end{tabular}
 }\\[1ex] 
 \tablefoot{
 See Table~\ref{t:list_g_v30e1}.
 }
 \end{table*}

%% file: tab_c3h7cn-n-g_v30e2_linelist.tex
\begin{table*}
 {\centering
 \caption{
 Selection of lines of \textit{gauche} propyl cyanide $\varv_{30}=2$ covered by the EMoCA survey of Sgr B2(N2).
}
 \label{t:list_g_v30e2}
 \vspace*{0.0ex}
 \begin{tabular}{lrcccrccrrcrr}
 \hline\hline
 \multicolumn{1}{c}{Transition\tablefootmark{a}} & \multicolumn{1}{c}{Frequency} & \multicolumn{1}{c}{Unc.\tablefootmark{b}} & \multicolumn{1}{c}{$E_{\rm up}$\tablefootmark{c}} & \multicolumn{1}{c}{$g_{\rm up}$\tablefootmark{d}} & \multicolumn{1}{c}{$A_{\rm ul}$\tablefootmark{e}} & \multicolumn{1}{c}{$\sigma$\tablefootmark{f}} & \multicolumn{1}{c}{$\tau_{\rm peak}$\tablefootmark{g}} & \multicolumn{2}{c}{Frequency range\tablefootmark{h}} & \multicolumn{1}{c}{$I_{\rm obs}$\tablefootmark{i}} & \multicolumn{1}{c}{$I_{\rm mod}$\tablefootmark{j}} & \multicolumn{1}{c}{$I_{\rm all}$\tablefootmark{k}} \\ 
  & \multicolumn{1}{c}{\scriptsize (MHz)} & \multicolumn{1}{c}{\scriptsize (kHz)} &  \multicolumn{1}{c}{\scriptsize (K)} & & \multicolumn{1}{c}{\scriptsize ($10^{-5}$ s$^{-1}$)} & \multicolumn{1}{c}{\scriptsize (mK)} & & \multicolumn{1}{c}{\scriptsize (MHz)} & \multicolumn{1}{c}{\scriptsize (MHz)} & \multicolumn{1}{c}{\scriptsize (K km s$^{-1}$)} & \multicolumn{2}{c}{\scriptsize (K km s$^{-1}$)} \\ 
 \hline
15$_{5,10}$ -- 14$_{5,9}$ &   90050.371 &   1 &  377 & 62 &  3.9 &  149 &  0.014 &  90049.2 &  90051.6 &   3.1(5)$^\star$ &   2.6 &   2.6 \\ 
16$_{6,11}$ -- 15$_{6,10}$ &   95820.263 &   1 &  385 & 66 &  4.6 &  100 &  0.015 &  95818.8 &  95821.8 &   4.8(4)$^\star$ &   3.0 &   3.1 \\ 
17$_{5,13}$ -- 16$_{5,12}$ &  102144.113 &   1 &  386 & 70 &  5.9 &  141 &  0.018 & 102142.7 & 102146.1 &   6.5(5)$^\star$ &   4.0 &   4.9 \\ 
17$_{3,14}$ -- 16$_{3,13}$ &  105441.876 &   1 &  382 & 70 &  6.9 &  123 &  0.021 & 105440.3 & 105443.7 &   3.8(5)$^\star$ &   3.5 &   3.7 \\ 
 \hline
 \end{tabular}
 }\\[1ex] 
 \tablefoot{
 See Table~\ref{t:list_g_v30e1}.
 }
 \end{table*}

%% file: tab_c3h7cn-n-a_v30e2_linelist.tex
\begin{table*}
 {\centering
 \caption{
 Selection of lines of \textit{anti} propyl cyanide $\varv_{30}=2$ covered by the EMoCA survey of Sgr B2(N2).
}
 \label{t:list_a_v30e2}
 \vspace*{0.0ex}
 \begin{tabular}{lrcccrccrrcrr}
 \hline\hline
 \multicolumn{1}{c}{Transition\tablefootmark{a}} & \multicolumn{1}{c}{Frequency} & \multicolumn{1}{c}{Unc.\tablefootmark{b}} & \multicolumn{1}{c}{$E_{\rm up}$\tablefootmark{c}} & \multicolumn{1}{c}{$g_{\rm up}$\tablefootmark{d}} & \multicolumn{1}{c}{$A_{\rm ul}$\tablefootmark{e}} & \multicolumn{1}{c}{$\sigma$\tablefootmark{f}} & \multicolumn{1}{c}{$\tau_{\rm peak}$\tablefootmark{g}} & \multicolumn{2}{c}{Frequency range\tablefootmark{h}} & \multicolumn{1}{c}{$I_{\rm obs}$\tablefootmark{i}} & \multicolumn{1}{c}{$I_{\rm mod}$\tablefootmark{j}} & \multicolumn{1}{c}{$I_{\rm all}$\tablefootmark{k}} \\ 
  & \multicolumn{1}{c}{\scriptsize (MHz)} & \multicolumn{1}{c}{\scriptsize (kHz)} &  \multicolumn{1}{c}{\scriptsize (K)} & & \multicolumn{1}{c}{\scriptsize ($10^{-5}$ s$^{-1}$)} & \multicolumn{1}{c}{\scriptsize (mK)} & & \multicolumn{1}{c}{\scriptsize (MHz)} & \multicolumn{1}{c}{\scriptsize (MHz)} & \multicolumn{1}{c}{\scriptsize (K km s$^{-1}$)} & \multicolumn{2}{c}{\scriptsize (K km s$^{-1}$)} \\ 
 \hline
21$_{1,21}$ -- 20$_{1,20}$ &   91576.003 &   2 &  343 & 43 &  6.9 &  117 &  0.021 &  91575.1 &  91577.5 &   4.0(4)$^\star$ &   2.8 &   2.9 \\ 
21$_{8,13}$ -- 20$_{8,12}$ &   93021.503 &   2 &  411 & 43 &  6.2 &  117 &  0.055 &  93020.3 &  93023.3 &   8.6(5) &   7.9 &  10.2 \\ 
21$_{8,14}$ -- 20$_{8,13}$ &   93021.503 &   2 &  411 & 43 &  6.2 & -- & -- & -- & -- & -- & -- & -- \\ 
21$_{5,17}$ -- 20$_{5,16}$ &   93022.018 &   2 &  369 & 43 &  6.9 & -- & -- & -- & -- & -- & -- & -- \\ 
21$_{5,16}$ -- 20$_{5,15}$ &   93022.027 &   2 &  369 & 43 &  6.9 & -- & -- & -- & -- & -- & -- & -- \\ 
23$_{0,23}$ -- 22$_{0,22}$ &  100963.335 &   2 &  352 & 47 &  9.3 &  141 &  0.025 & 100962.2 & 100965.1 &   6.7(5)$^\star$ &   5.2 &   5.5 \\ 
 \hline
 \end{tabular}
 }\\[1ex] 
 \tablefoot{
 See Table~\ref{t:list_g_v30e1}.
 }
 \end{table*}

%% file: tab_c3h7cn-n-g_v28e1_linelist.tex
\begin{table*}
 {\centering
 \caption{
 Selection of lines of \textit{gauche} propyl cyanide $\varv_{28}=1$ covered by the EMoCA survey of Sgr B2(N2).
}
 \label{t:list_g_v28e1}
 \vspace*{0.0ex}
 \begin{tabular}{lrcccrccrrcrr}
 \hline\hline
 \multicolumn{1}{c}{Transition\tablefootmark{a}} & \multicolumn{1}{c}{Frequency} & \multicolumn{1}{c}{Unc.\tablefootmark{b}} & \multicolumn{1}{c}{$E_{\rm up}$\tablefootmark{c}} & \multicolumn{1}{c}{$g_{\rm up}$\tablefootmark{d}} & \multicolumn{1}{c}{$A_{\rm ul}$\tablefootmark{e}} & \multicolumn{1}{c}{$\sigma$\tablefootmark{f}} & \multicolumn{1}{c}{$\tau_{\rm peak}$\tablefootmark{g}} & \multicolumn{2}{c}{Frequency range\tablefootmark{h}} & \multicolumn{1}{c}{$I_{\rm obs}$\tablefootmark{i}} & \multicolumn{1}{c}{$I_{\rm mod}$\tablefootmark{j}} & \multicolumn{1}{c}{$I_{\rm all}$\tablefootmark{k}} \\ 
  & \multicolumn{1}{c}{\scriptsize (MHz)} & \multicolumn{1}{c}{\scriptsize (kHz)} &  \multicolumn{1}{c}{\scriptsize (K)} & & \multicolumn{1}{c}{\scriptsize ($10^{-5}$ s$^{-1}$)} & \multicolumn{1}{c}{\scriptsize (mK)} & & \multicolumn{1}{c}{\scriptsize (MHz)} & \multicolumn{1}{c}{\scriptsize (MHz)} & \multicolumn{1}{c}{\scriptsize (K km s$^{-1}$)} & \multicolumn{2}{c}{\scriptsize (K km s$^{-1}$)} \\ 
 \hline
15$_{7,9}$ -- 14$_{7,8}$ &   90093.047 &   5 &  428 & 62 &  3.4 &  149 &  0.018 &  90091.7 &  90094.6 &   4.1(6)$^\star$ &   3.4 &   4.5 \\ 
15$_{7,8}$ -- 14$_{7,7}$ &   90093.194 &   5 &  428 & 62 &  3.4 & -- & -- & -- & -- & -- & -- & -- \\ 
 \hline
 \end{tabular}
 }\\[1ex] 
 \tablefoot{
 See Table~\ref{t:list_g_v30e1}.
 }
 \end{table*}

%% file: tab_c3h7cn-n-a_v29e1_linelist.tex
\begin{table*}
 {\centering
 \caption{
 Selection of lines of \textit{anti} propyl cyanide $\varv_{29}=1$ covered by the EMoCA survey of Sgr B2(N2).
}
 \label{t:list_a_v29e1}
 \vspace*{0.0ex}
 \begin{tabular}{lrcccrccrrcrr}
 \hline\hline
 \multicolumn{1}{c}{Transition\tablefootmark{a}} & \multicolumn{1}{c}{Frequency} & \multicolumn{1}{c}{Unc.\tablefootmark{b}} & \multicolumn{1}{c}{$E_{\rm up}$\tablefootmark{c}} & \multicolumn{1}{c}{$g_{\rm up}$\tablefootmark{d}} & \multicolumn{1}{c}{$A_{\rm ul}$\tablefootmark{e}} & \multicolumn{1}{c}{$\sigma$\tablefootmark{f}} & \multicolumn{1}{c}{$\tau_{\rm peak}$\tablefootmark{g}} & \multicolumn{2}{c}{Frequency range\tablefootmark{h}} & \multicolumn{1}{c}{$I_{\rm obs}$\tablefootmark{i}} & \multicolumn{1}{c}{$I_{\rm mod}$\tablefootmark{j}} & \multicolumn{1}{c}{$I_{\rm all}$\tablefootmark{k}} \\ 
  & \multicolumn{1}{c}{\scriptsize (MHz)} & \multicolumn{1}{c}{\scriptsize (kHz)} &  \multicolumn{1}{c}{\scriptsize (K)} & & \multicolumn{1}{c}{\scriptsize ($10^{-5}$ s$^{-1}$)} & \multicolumn{1}{c}{\scriptsize (mK)} & & \multicolumn{1}{c}{\scriptsize (MHz)} & \multicolumn{1}{c}{\scriptsize (MHz)} & \multicolumn{1}{c}{\scriptsize (K km s$^{-1}$)} & \multicolumn{2}{c}{\scriptsize (K km s$^{-1}$)} \\ 
 \hline
21$_{8,13}$ -- 20$_{8,12}$ &   92790.120 &   1 &  519 & 43 &  6.2 &  117 &  0.026 &  92789.3 &  92792.2 &   4.8(5)$^\star$ &   3.8 &   4.0 \\ 
21$_{8,14}$ -- 20$_{8,13}$ &   92790.120 &   1 &  519 & 43 &  6.2 & -- & -- & -- & -- & -- & -- & -- \\ 
21$_{5,17}$ -- 20$_{5,16}$ &   92790.660 &   1 &  479 & 43 &  6.8 & -- & -- & -- & -- & -- & -- & -- \\ 
21$_{5,16}$ -- 20$_{5,15}$ &   92790.670 &   1 &  479 & 43 &  6.8 & -- & -- & -- & -- & -- & -- & -- \\ 
 \hline
 \end{tabular}
 }\\[1ex] 
 \tablefoot{
 See Table~\ref{t:list_g_v30e1}.
 }
 \end{table*}